\documentclass{emulateapj}
\usepackage{apjfonts}
 \slugcomment{To appear in ApJ, 10 Apr 2007, vol 659} 
\shorttitle{Stellar Variability in Galactic Center}
\shortauthors{Rafelski, Ghez, Hornstein, Lu, \& Morris}
 
\begin{document}

\title{Photometric Stellar Variability in the Galactic Center}

\author{M. Rafelski, A. M. Ghez\altaffilmark{1}, S. D. Hornstein, J. R. Lu, M. Morris}
\affil{Division of Astronomy and Astrophysics, UCLA, Los Angeles, CA 90095-1547}
\email{marcar, ghez, seth,  jlu, morris@astro.ucla.edu}
\altaffiltext{1}{Institute of Geophysics and Planetary Physics, University of 
California, Los Angeles, CA 90095-1565}

\begin{abstract}

We report the results of a diffraction-limited, photometric variability 
study of  the central 5$\tt''$ $\times$ 5$\tt''$ of the Galaxy conducted over the past 10 years
using speckle imaging techniques on the W. M. Keck I 10 m telescope.
Within our limiting magnitude of $m_{K} < 16$ mag for images made from a single
night of data, we find a minimum of 15 K[2.2 $\micron$]-band variable stars out of 131 monitored stars.
While large populations of binaries have been posited to exist in
this region, both to explain the presence of young stars in the vicinity of
a black hole and because of the high stellar densities, only two binaries are identified 
in this study. First is the previously identified Ofpe/WN9 equal mass 
eclipsing binary star IRS 16SW, for which we measure an orbital period
of $19.448 \pm 0.002$ days.
In contrast to recent results, our data on IRS 16SW 
show an asymmetric phased light curve with
a much steeper fall-time than rise-time,
which may be due to tidal deformations caused
by the proximity of the stars in their orbits. 
Second is the WC 9 Wolf-Rayet star IRS 29N; its observed photometric variation over a 
few year time-scale is likely due to episodic dust production in a 
binary system containing two windy stars.
Our sample also includes 4 candidate
Luminous Blue Variable (LBV) stars (IRS 16NE, 16C, 16NW, 16SW).
While 2 of them show variability, none show the  
characteristic of LBVs large increase or decrease in luminosity. However, our  
time baseline is too short to rule them out as LBVs. Nonetheless, the lack of
evidence for these stars to be LBVs and their coexistence with a significant surrounding population 
of well established Wolf-Rayet stars is consistent with needing only a single recent starburst
event at the Galactic center to account for all of the known, young, massive stars. 
Among the remaining variable stars, the majority are early-type stars and three are 
possibly variable due to line of sight extinction variations. 
For the 7 OB stars at the center of our field of view that have well-determined 3-dimensional 
orbits, we see no evidence of flares or dimming of their light, which
limits the possibility of a cold, geometrically-thin inactive accretion disk
around the supermassive black hole, Sgr A$^{*}$.

\end{abstract}


\pagebreak
\clearpage

\section{Introduction}

The stellar cluster at the Galactic center (GC) presents a unique opportunity 
to study the evolution and properties of stars within the sphere of influence of
a $3-4\times 10^{6}M_{\odot}$ supermassive black hole (SMBH) \citep{ghez03, ghez05a, sch03}. 
Photometric variability offers a useful approach to a number of outstanding questions regarding
this stellar population which is composed of a mixture of old giants and young, massive stars.
\citep{krab91, krab95, blum96a, blum96b, blum03, fig03, pau01, pau04b, pau06}. 
For example, light curves can easily reveal
close binary stars, which are relevant in several ways to
our understanding of stars at the Galactic center.  

First, binaries on radial orbits that are disrupted by the central black hole
may provide a mechanism for capturing young stars from large
galacto-centric radii, where the conditions are conducive to
star formation, and retaining them at the smaller less hospitable radii where
many young stars are found today \citep{andy03}.
Second, binary companions may facilitate the production of dust around the WC sub-class
of Wolf-Rayet stars, which are massive post-main sequence stars undergoing
rapid mass loss.  While conditions in the hostile environment (high 
temperatures in particular) of the stellar winds do not favor 
the formation of dust \citep{williams87}, compression within 
wind-colliding binary systems could overcome this challenge 
\citep{white95, veen98, williams00, lef05}.	
Third, binaries provide 
a direct measurement of stellar masses.  
This is especially helpful for the most massive 
stars in the Galactic center as 
it would assist our understanding of the 
recent star formation history.

Another way in which a photometric variability study constrains the
recent star formation history, as well as our understanding of massive
star evolution, is the possibility of identifying luminous blue variables
(LBVs).  There are currently only 12 confirmed Galactic LBVs and 23 additional
candidates, with 6 candidates in the Galactic center IRS 16 cluster of stars
alone \citep{clark05}.  The LBV phase plays an important, although
poorly constrained, role in stellar
evolution, because, during this phase, stars experience significant
mass loss, with rates of $\sim10^{-2}M_{\odot}$/yr during eruptions and as
high as high as $10^{-4.5}M_{\odot}$/yr during quiescent phases
\citep{abb87, hum94, mas03}.
From a star formation history stand point,
the LBV phase is notable because it is the first of several post main sequence
phases that only the most massive stars ($M \gtrsim 60-85 M_{\odot}$) may go
through before becoming supernovae.   Stars stay in this phase for
only $\sim 10^{4}$ years
\citep{stoth96} before entering the Wolf-Rayet phase, which
typically lasts a few $\times 10^{6}$ years \citep{mm05}.
Less massive stars ($M \gtrsim40 M_{\odot}$) will skip the LBV phase and become
Wolf-Rayet stars, but on time-scales longer than that of the more massive
stars that experienced an LBV phase.  Therefore, in principle, the numbers of
LBVs and WR stars can constrain recent star formation
histories (e.g. \citealt{pau06, fig04}). In this context
the candidate LBVs at the Galactic center are perplexing in the context of
the $\gtrsim 25$ Wolf-Rayet stars located in their immediate vicinity,
since in a single starburst event one would not expect to see any WR stars if
the most massive stars are just now evolving through the LBV phase.
This is similar to the problem posed by the presence of two LBVs in the
Quintuplet cluster \citet{fig04}.
If confirmed, the LBV candidates would suggest that this region has undergone
multiple recent star forming events or that our understanding of LBV evolution is incomplete.

The photometry of stars in close proximity to the SMBH can also be used to constrain
the properties of a possible cold, geometrically-thin inactive accretion disk around
Sgr A$^{*}$ which could explain the
present-day low luminosity of Sgr A$^{*}$ \citep{nay03, cua03}.
In the presence of such a disk, we would expect to see nearby stars
eclipsed or reddened when they pass behind the disk.

Very few photometric variable studies of the Galactic center exist. 
\citet{tam96} introduced the idea that stars close to the Galactic center 
are expected to have a higher fraction of ellipsoidal\footnote{Ellipsoidal variables
are non-eclipsing binaries that are elongated by mutual tidal forces \citep{sterk96}.} and eclipsing 
variable binaries than the stars in the solar neighborhood, but found very 
few variable stars and no binary stars. 
Seeing-limited studies \citep{tam96, blum96a} are limited to the brightest stars, 
due to stellar confusion caused by the high stellar densities and proper motions 
close to the central black hole.  With high angular resolution data, \citet{ott99} 
have identified the only known eclipsing binary system in this region 
(see also \citealt{depoy04, mar06}).  
Furthermore, they suggest that as much as half of their sample (K $\lesssim$13) may be variable.
However, the variability fraction decreases at smaller galactocentric radii starting from $\sim5\tt''$, 
suggesting that even at a high resolution of 0$\farcs$13 their sensitivity to 
variability is limited by stellar confusion.

In this paper, we present the results of a stellar variability study of the
central 5$\tt''$$\times$5$\tt''$ of our Galaxy, based on ten years of K[2.2 $\mu$m] diffraction-limited
images from the W. M. Keck I Telescope ($\theta=0\farcs05$). 
The observations are described in \S2, and
the data and methodology to determine variability in \S3.  We discuss the
variable star population in \S4, which includes identification of asymmetries
in the light curve of the eclipsing binary star IRS 16SW and the discovery 
of a likely wind colliding binary star in IRS 29N, and summarize our major findings in \S5.

\section{Observations}

K-band ($\lambda_{o}$ = 2.2\micron, $\Delta\lambda$=0.4\micron) speckle imaging observations 
of the Galaxy's central stellar cluster were obtained with the W. M. Keck I 10 m telescope using the 
facility near-infrared camera, NIRC \citep{matt94}. 
Observations taken from 1995 to 2004 have been described
in detail elsewhere \citep{ghez98, ghez00, ghez05a, lu05} and new
observations on 2005 April 24-25 were conducted in a similar manner,
resulting in diffraction-limited images.
Each night several thousand short-exposure frames were taken in sets of $\sim200$, 
with NIRC in its fine plate scale mode, which has a scale of  20.40 
$\pm$ 0.04 mas pixel$^{-1}$ and a corresponding 
field of view (FOV) of 5$\farcs$22 $\times$ 5$\farcs$22 \citep{matt96}. 
Table \ref{tab1}  lists the date and number of frames obtained for
each of the 50 nights of observations used in this study.

\vspace{0.4in}

\section{Data Analysis and Results}

\subsection{Image Processing}

The individual frames are processed in two steps to create a final average
image for each night of observation.  First, the standard image
reduction steps of sky subtraction, flat-fielding, bad pixel
correction, optical distortion correction\footnote{
http://www.keck.hawaii.edu/inst/nirc/Distortion.html},
and pixel magnification by a factor of two are carried out on
each frame.  Second, the frames from each night of observation
are combined using the method of "Shift-and-Add" \citep{chris91}
with the frame selection and weighting scheme prescribed by
\citet{horn06}.  In short, each frame is analyzed for
Strehl quality using the peak pixel value of IRS 16C, and low quality
frames, which do not improve the cumulative signal to noise ratio
(SNR) for the observations from each night, are rejected.  This typically
leaves $\sim$1600 frames for each night or 37\% of the
original data set (see column 3 of Table \ref{tab1}).  The remaining frames
for a given night are combined with
Shift-and-Add in an average that is weighted by each frame's
peak pixel value for IRS 16C.  The final images have typical Strehl ratios
of $\sim$0.07 (see column 8 of Table \ref{tab1}).
The dataset from each night is also divided into three 
equivalent quality (and randomized in time) subsets
to make three independent weighted Shift-and-Add  image subsets, which
are used to determine measurement uncertainties and to reject
spurious sources.

\begin{deluxetable*}{lrrccrc}
\tabletypesize{\scriptsize}
\tablecaption{List of Observations
\label{tab1}}
\tablewidth{0pt}
\tablehead{
\colhead{Date} & 
\colhead{Frames\tablenotemark{a} }   & 
\colhead{Frames\tablenotemark{b}} &
\colhead{Num. Stars\tablenotemark{c}} &
\colhead{Num. Stars\tablenotemark{d}} &
\colhead{SNR\tablenotemark{e}} & 
\colhead{Strehl} \\
\colhead{} &
\colhead{(Obs.)} &
\colhead{(Used) } &
\colhead{(Initial)} &
\colhead{(Final)} &
\colhead{} &
\colhead{}
}

\startdata

1995 Jun 10 &1200 &425 &54 &66 &10.3 &0.08 \\
1995 Jun 11 &2700 &1604 &95 &110 &15.0 &0.06 \\
1995 Jun 12 &2100 &1082 &107 &108 &12.0 &0.04 \\
1996 Jun 26 &4200 &585 &116 &119 &19.8 &0.04 \\
1996 Jun 27 &2300 &1260 &117 &121 &20.5 &0.04 \\
1997 May 14 &3600 &1851 &63 &82 &16.7 &0.06 \\
1998 Apr 02 &2660 &1649 &119 &121 &16.3 &0.05 \\
1998 May 14 &4560 &1748 &96 &114 &16.7 &0.04 \\
1998 May 15 &7030 &1953 &41 &55 &9.9 &0.06 \\
1998 Jul 04 &2280 &943 &108 &114 &17.8 &0.08 \\
1998 Aug 04 &6270 &1469 &87 &107 &15.5 &0.05 \\
1998 Aug 05 &5700 &1592 &94 &103 &17.3 &0.07 \\
1998 Oct 09 &2660 &1188 &83 &99 &17.1 &0.08 \\
1998 Oct 11 &570 &450 &79 &96 &15.0 &0.05 \\
1999 May 02 &7030 &1589 &115 &116 &19.9 &0.09 \\
1999 May 03 &2090 &1264 &103 &114 &16.0 &0.07 \\
1999 Jul 24 &5510 &2239 &113 &119 &18.5 &0.11 \\
1999 Jul 25 &950 &788 &102 &114 &15.0 &0.06 \\
2000 Apr 21 &3040 &947 &90 &112 &18.5 &0.04 \\
2000 May 19 &9880 &1970 &64 &91 &10.2 &0.09 \\
2000 May 20 &7600 &2146 &81 &97 &16.5 &0.10 \\
2000 Jul 19 &8740 &1939 &111 &116 &17.1 &0.07 \\
2000 Jul 20 &3420 &1454 &111 &119 &19.8 &0.09 \\
2000 Oct 18 &2280 &1807 &82 &107 &15.1 &0.05 \\
2001 May 08 &1520 &889 &111 &124 &20.1 &0.04 \\
2001 May 09 &6270 &1990 &76 &100 &13.1 &0.08 \\
2001 Jul 28 &4180 &1752 &99 &104 &20.0 &0.13 \\
2001 Jul 29 &6080 &1751 &105 &115 &19.0 &0.07 \\
2002 Apr 23 &7410 &1669 &117 &119 &17.2 &0.05 \\
2002 Apr 24 &7790 &1882 &104 &113 &19.6 &0.06 \\
2002 May 23 &1900 &1249 &66 &84 &18.3 &0.07 \\
2002 May 24 &2660 &1537 &85 &100 &17.7 &0.09 \\
2002 May 28 &2850 &1866 &59 &80 &13.2 &0.06 \\
2002 May 29 &3420 &1552 &86 &104 &16.8 &0.07 \\
2002 Jun 01 &5510 &1992 &43 &53 &7.7 &0.09 \\
2002 Jul 19 &4370 &1115 &41 &57 &7.0 &0.07 \\
2002 Jul 20 &3990 &1355 &51 &64 &11.0 &0.06 \\
2003 Apr 21 &5130 &1799 &93 &107 &19.1 &0.04 \\
2003 Apr 23 &5320 &1970 &69 &87 &13.2 &0.05 \\
2003 Jul 22 &5130 &1718 &70 &94 &16.6 &0.08 \\
2003 Sep 07 &4560 &1795 &108 &112 &16.3 &0.07 \\
2003 Sep 08 &4370 &1223 &97 &110 &12.3 &0.07 \\
2004 Apr 29 &6840 &1181 &53 &68 &14.5 &0.11 \\
2004 Apr 30 &4180 &1203 &98 &105 &16.6 &0.05 \\
2004 Jul 25 &5320 &2007 &98 &110 &18.1 &0.08 \\
2004 Jul 26 &8550 &2309 &33 &38 &6.7 &0.08 \\
2004 Aug 29 &3230 &1328 &120 &122 &21.0 &0.10 \\
2005 Apr 24 &7410 &2195 &51 &60 &11.6 &0.07 \\
2005 Apr 25 &9500 &2035 &116 &116 &20.8 &0.05 \\
2005 Jul 26 &6650 &1497 &98 &113 &19.0 &0.06 \\

\enddata

\tablecomments{All observations are speckle K-band 
($\lambda_{o}$ = 2.2\micron, $\Delta\lambda$=0.4\micron) images.}

\tablenotetext{a}{The number of frames observed in the night in stacks of 190 frames.}
\tablenotetext{b}{The number of frames used in weighted shift-and-add routine described in \cite{horn06}.}
\tablenotetext{c}{Number of stars in initial source list.}
\tablenotetext{d}{Number of stars in final source list.}
\tablenotetext{e}{The signal to noise ratio determined from median uncertainties of the 
six faintest non-variable stars detected in all the nights (S0-14, S1-25, S0-13, S1-68, S2-5, S1-34)
with $m_{k} \sim 13.4$ mag.}

\end{deluxetable*}


Sources are identified in individual images and cross-identified
between images using the strategy developed in \citet{ghez98, ghez00, ghez05a} 
and \citet{lu06}, which for this study entails four separate steps.
In the first step of the source identification process,
we generate a conservative initial list of sources for each 
night of data to help minimize spurious source detection. 
This is done using the point-spread-function (PSF) fitting routine
 {\it StarFinder} \citep{dio00} to identify sources in both the average images and the subset images.
StarFinder identifies sources through cross-correlation of each image with its PSF model, which, 
for our implementation, is generated from the two bright stars IRS 16C and IRS 16NW.
The initial source list for each night of data is composed of 
only sources detected in the average images with correlation values above 0.8 and
in all three subset images with correlation values above 0.6. 
In the second step of the source identification process, the source lists from all nights are 
cross-identified to produce a master list of sources, using a process that is 
described in \citet{ghez98} and that also solves for the sources' proper motions.
To further ensure that no spurious sources have been detected 
we require that sources be detected in a minimum of 13 nights\footnote{The threshold for
the minimum number of nights was chosen by looking 
for a drop in the distribution of the number of nights
that the sources were detected in the first pass at source identification. A minor drop is seen at 
13 nights. The final results are not very sensitive to this choice and we 
therefore have made a fairly conservative choice.}. 

\begin{figure*}
\plotone{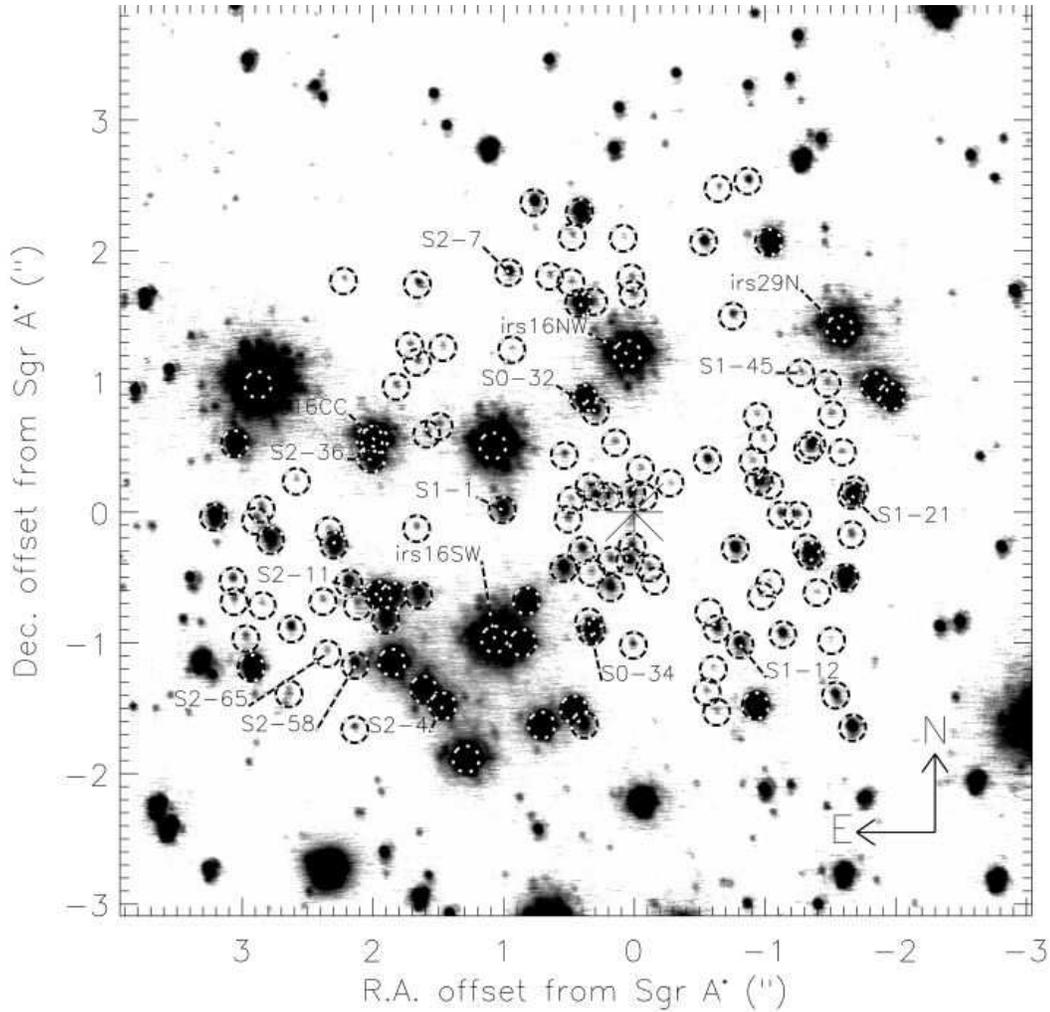}
\caption{ 
Identification of sources from this study overlaid on 
a 6$\tt''$ x 6$\tt''$ region of an LGS K$\tt'$-band ($\lambda_{o}$ = 2.1\micron)
image from \citet{ghez06} taken on June 30, 2005.
All 131 sources are circled, but only variable sources are labeled.
The location of Sgr A$^{*}$ is marked with an asterisk.
\label{sources}}
\end{figure*}

\begin{figure}
\plotone{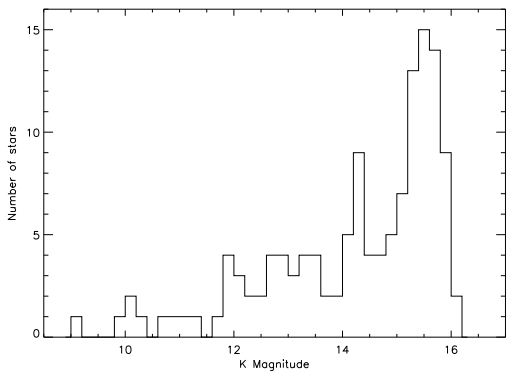}
\caption{A $m_{K}$ histogram of the 131 stars in our sample which range in magnitude from 9.0 to 16.1 mag.
\label{histmag}}
\end{figure}

Figure \ref{sources} displays the 131 sources contained in our final master list.
In the third step of the source identification process, 
we return to the original images to search more aggressively
for the sources on the master list that were missed in some of the images.  
We explicitly feed the master list of sources into StarFinder and search for only these
sources at their predicted positions with more lenient criteria, which require
correlation values above 0.4 for both average and subset images.
In the fourth and final step, we impose a restriction on our
source detections to ensure photometric reliability: we
exclude source detections that occur in regions of the average images 
covered by less than 50\% frames that went into making a
particular image. These regions, which are on the edges of the image, 
have relatively low signal to noise and the PSFs in these regions may not 
be well represented by the PSF model. We also exclude individual 
measurements in which known stars are blended with each other 
(i.e., sources as listed in \citet{ghez05a} as well as Sgr A$^*$ IR.)
This procedure, in its entirety, produces 4795 detections among 131 sources, 
which range in $m_{K}$ magnitude from 9.0 to 16.1 mag (see Figure \ref{histmag}).


\begin{figure}
\plotone{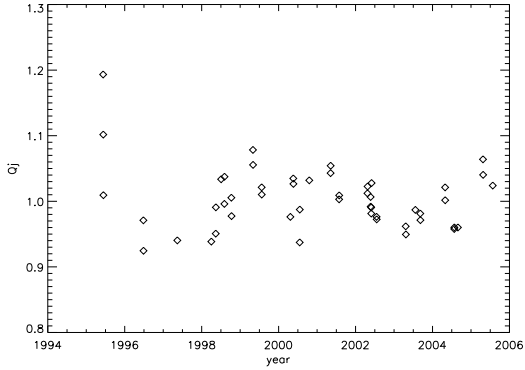}
\caption{The average normalized flux densities ($Q_{j}$, as defined in \S 3.1)
as a function of time. The variations seen are consistent with the measurement uncertainties for 
bright stars and are therefore demonstrative of the stability of our calibrator IRS 16C. 
This quantity is used as a scale factor to improve the quality of our relative photometry. 
\label{sys}}
\end{figure}

\begin{figure}
\plotone{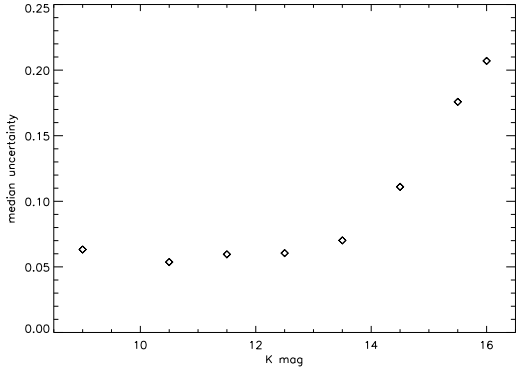}
\caption{The median measurement uncertainty of stars binned by magnitude. 
The median uncertainties grow from a floor of about 0.06 mag to 0.21 mag for the K= 16 mag sources.
\label{bias}}
\end{figure}

Photometric zeropoints are established on the basis of the work done by \citet{blum96a}.
While we share 7 stars in common with \citet{blum96a}
(IRS 16NW, 16SW, 16C, 16NE, 29N, 29S, and 16CC), only IRS 16C 
($m_{K}=9.83 \pm 0.05$ mag) is a suitable photometric reference source.  
IRS 16SW is a known variable star in the K-band \citep{ott99, depoy04}
and IRS 29N is noted as possibly variable in \citet{horn02}. 
IRS 16CC appears to be variable in the L-band;
\citet{blum96a} list a re-calibrated value from \citet{depoy91} 
of $8.7 \pm 0.2$ mag, while \citet{sim96} measure $10.2 \pm 0.2$ mag.
Among the remaining sources, only IRS 16C is in the final source lists of all the images.
Several non-variable sources (see \S 3.2.1) are used {\it a posteriori} to confirm that 
IRS 16C is non-varying. Specifically, 
we check for any systematic shifts in the zero points by examining the
normalized flux densities
($Q_{j}=\sum_{i}^{N}\frac{Flux_{i, j}}{NFlux_{avg_{i}}}$
where index i represents each star in an image of epoch j, 
$Flux_{avg_{i}}$ is the weighted average of the flux for that 
star over all images, and N is the number of stars used)
of the 7 least variable bright stars that are identified in all 50 images
(S1-3, S1-5, S2-22, S2-5, S1-68, S0-13, and S1-25) (see \S 3.2.1). 
The photometric stability of IRS 16C is 
shown in Figure \ref{sys}, which
plots $Q_{j}$ versus the observing dates. 
The reference source IRS 16C appears to be stable over time, since
the standard deviation of $Q_{j}$ is 0.05, 
which is consistent with our measurement uncertainty for bright stars. 
Increasing the number of reference stars to 11 non-variable sources present
in all frames in all 50 nights yields the same result. 
We therefore conclude that IRS 16C is non-varying to within our measurement uncertainties,
and include it in our list of non-varying sources.
Uncertainty in each of our reported relative
photometry values is initially estimated
as the root mean square (RMS) deviation from the average of the
measurements from the three different subset images. The RMS value is 
added in quadrature with the uncertainty in the brightness of IRS 16C (0.05 mag)
determined from the standard deviation of the normalized flux densities $Q_{j}$.
As Figure \ref{bias} shows, the median uncertainties grow from a floor
of about 0.06 mag to 0.21 mag for the K= 16 mag sources. 

\vspace{0.5in}

\subsection{Variability}

\subsubsection{Identifying Variables}

\begin{figure}
\plotone{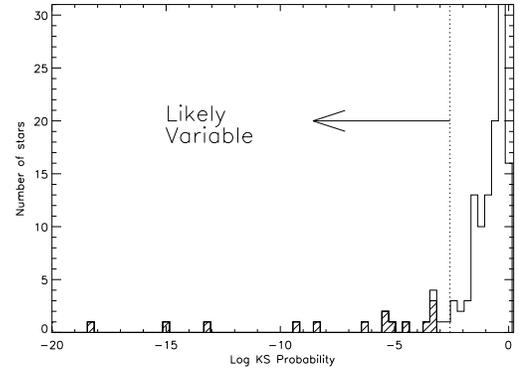}
\caption{KS test probabilities of all the stars, with low values implying a low probability for a star to be 
non-variable. We classify a star as variable if it has a KS probability of less than $2.7 \times 10^{-3}$,
which is equivalent to a  $3\sigma$ cutoff value for gaussian distributed uncertainties and is marked
by the dashed line. To ensure all our low KS probability stars are truly variable, we require all variable 
stars to have positive intrinsic variance. We mark all variable stars identified in this survey with hatches.
\label{ksprob}}
\end{figure}

There is a wide range of methods for testing photometric variability
and the challenge for these various approaches is to avoid
declaring a non-variable source variable on the basis of a few outlying
data points \citep{welch93}. We therefore have
chosen to use the Kolmogorov-Smirnov (KS) test to
calculate the probability that a {\it distribution} of 
data points is consistent with a model of a distribution 
of measurements for a non-variable source.
This approach is less sensitive to outlying data points 
than the commonly used $\chi^2$ test, which is an analysis of a 
single number description of how well a
data set matches a model.
In the KS test, we adopt as our model a non-variable light curve with
gaussian-distributed uncertainties and we test the consistency of the measurements 
with the model. Specifically, we examine the distribution of 
$X_{j} = \frac{Flux_{j}-Flux_{avg}}{\sigma_{Flux_{j}}}$ 
where $Flux_{j}$ is the flux of a star in a image of epoch j,
$\sigma_{Flux_{j}}$ is the corresponding uncertainty,
and  $Flux_{avg}$ is the weighted average of the flux for that star over all images.
The resulting KS probabilities, which have allowed values between 0 and
1, describe how likely it is that a source's measurements are consistent with a non-variable
source.  Therefore variable stars, whose intensity variations are 
larger or comparable to our measurement uncertainties, 
should have very low KS probabilities.
We classify a star as variable if it has a KS probability of less than $2.7 \times 10^{-3}$,
which is the equivalent to a $3\sigma$ cut for gaussian distributed uncertainties
(see Figure \ref{ksprob}).
To ensure all our low KS probability stars are truly variable, we 
also require these stars to have positive estimates of their intrinsic flux density variance,
$\sigma^{2}_{intrinsic} = \sigma^{2}_{measured}-\langle\delta\rangle^{2}$, 
where the first term is the dispersion of the measured flux densities and the second term removes
the bias introduced by the measurement uncertainties, $\delta$.
At this point, the 7 least variable bright stars detected in all images,
used in \S 3.1 to define $Q_{j}$, are identified.
We then scale all our photometry by $Q_{j}$ in order to reduce
the fluctuations induced by measurement errors on IRS 16C.
The KS and intrinsic variance tests are then repeated.
Table \ref{tab2} and \ref{tab3} list the properties of the variable and non-variable 
stars in our sample, respectively, and the light curves of all variable stars and a few 
key non-variable stars are shown below in Figures \ref{lclbv}, \ref{lcwc}, \ref{lcob}, 
\ref{lcearly}, \ref{lcmisc}, and \ref{lcorbit}.
While there are almost certainly 
other variable stars that we have excluded,
our uncertainties limit our ability to classify more of these stars as variable, 
especially at the fainter end. 

Among the 131 stars in our sample, 15 are identified as photometric
variables in K-band (see Figure \ref{sources}). To this we also add IRS 16CC,
known to be variable in the L-band (see \S 3.1).  
Since the relative photometric uncertainties 
are roughly uniform down to a $m_{K}$ magnitude 
of $\sim$ 14 and then grow at fainter magnitudes (see Figure \ref{bias}), we report
a frequency of variable stars based on the stars brighter
than 14 mag.  Within this brighter sample of 44 stars, there are
10 variable stars, suggesting a minimum frequency of variable stars
of 23\%. 
There is no evidence for radial dependence, suggesting that we 
are not limited by stellar confusion down to 14 mag.

We compare our results to those of \citet{ott99}, the only other high
spatial resolution study of the variability of sources in the Galactic center.
Those authors give an upper limit of possible variable stars of approximately
50\% of their 218 sources with $m_{K}$ $<$ 13 mag over 18$\tt''$ $\times$ 18$\tt''$.
While this variable star frequency is higher than our reported value (and consistent),
a comparison limited to the stars in common leads to a number of discrepancies.
In the overlap sample of 33 stars, \citeauthor{ott99}
find 2 of the stars to be variable (IRS 16SW, S1-3), while our sample has 6 
(IRS 16SW, IRS 16NW, IRS 29N, S2-11, S2-4, S1-21) and only IRS 16SW is in common. 
There are a number of differences between these two studies, including the data analysis
approach used (PSF fitting vs.
\begin{deluxetable*}{lrcccrrccr}
\tabletypesize{\scriptsize}
\tablecaption{List of Variable Stars
\label{tab2}}
\tablewidth{0pt}
\tablehead{
\colhead{Star ID} &
\colhead{Other ID} &  
\colhead{K\tablenotemark{a}}   & 
\colhead{Int. Var.} &
\colhead{p}   & 
\colhead{$\Delta$R.A.} &
\colhead{$\Delta$Dec.} &
\colhead{Probability} & 
\colhead{Nights} &
\colhead{Type} \\
\colhead{} &
\colhead{} &
\colhead{(mag)} &
\colhead{(mag)} &
\colhead{(arcsec)} &
\colhead{(arcsec)} &
\colhead{(arcsec)} &
\colhead{} &
\colhead{(days)} &
\colhead{}
}

\startdata

IRS16SW &E23 &10.06$ \pm $0.19 &0.17 &1.41 &1.04 &-0.95 &2.3E-23 &50 &Ofpe/WN9\tablenotemark{c}\tablenotemark{h} \\
IRS16NW &E19 &10.09$ \pm $0.09 &0.07 &1.21 &-0.01 &1.21 &3.3E-05 &49 &Ofpe/WN9\tablenotemark{c}\tablenotemark{g} \\
IRS29N &E31 &10.33$ \pm $0.20 &0.18 &2.15 &-1.63 &1.40 &1.1E-15 &35 &WC9\tablenotemark{c}\tablenotemark{g} \\
IRS16CC &E27 &10.60$ \pm $0.05 &0.01 &2.07 &1.99 &0.57 &3.0E-01 &50 &O9.5-B0.5 I\tablenotemark{c} \tablenotemark{i} \\
S2-11 &GEN+2.03-0.63 &11.99$ \pm $0.13 &0.11 &2.07 &1.99 &-0.58 &4.9E-19 &49 &Late\tablenotemark{e}\tablenotemark{g} \\
S2-4 &E28:GEN+1.46-1.49 &12.26$ \pm $0.17 &0.15 &2.05 &1.45 &-1.45 &6.1E-14 &47 &B0-0.5 I\tablenotemark{c}\tablenotemark{g} \\
S1-1 &GEN+1.01+0.02 &13.00$ \pm $0.11 &0.08 &0.98 &0.98 &0.05 &3.6E-04 &49 &Early\tablenotemark{f} \\
S2-36 &\nodata &13.28$ \pm $0.13 &0.12 &2.08 &2.04 &0.43 &3.7E-09 &48 &Early\tablenotemark{f} \\
S1-21 &E24:W7 &13.33$ \pm $0.17 &0.15 &1.69 &-1.69 &0.13 &4.2E-06 &42 &O9-9.5 III?\tablenotemark{c}\tablenotemark{g} \\
S1-12 &E21:W13 &13.82$ \pm $0.18 &0.17 &1.31 &-0.85 &-1.00 &4.2E-07 &45 &OB I?\tablenotemark{c} \\
S2-7 &E29:GEN+1.06+1.81 &14.06$ \pm $0.25 &0.21 &2.09 &0.97 &1.85 &4.2E-10 &45 &O9-B0\tablenotemark{c} \\
S0-32 &\nodata &14.18$ \pm $0.20 &0.15 &0.81 &0.26 &0.77 &6.3E-04 &49 &Early\tablenotemark{f} \\
S2-58 &\nodata &14.21$ \pm $0.14 &0.10 &2.45 &2.17 &-1.14 &6.8E-04 &47 &Early\tablenotemark{f} \\
S1-45 &\nodata &15.41$ \pm $0.55 &0.28 &1.63 &-1.28 &1.00 &6.1E-06 &41 &Early\tablenotemark{f} \\
S2-65 &\nodata &15.83$ \pm $0.49 &0.29 &2.57 &2.37 &-1.00 &2.5E-04 &29 &\nodata \\
S0-34 &\nodata &15.85$ \pm $0.40 &0.31 &0.83 &0.32 &-0.77 &4.1E-06 &26 &\nodata \\

\enddata

\tablecomments{Photometry is relative to IRS 16C ($m_{k}$=9.83 mag). 
Positions are in arcseconds offset from Sgr A* in 1999.56 and p is the projected distance.
The K-S probability is equal to 1 for an ideal non variable 
source, and approaches 0 for a very variable source. 
Other ID's are from \citet{pau06} and \citet{gen00} respectively.
We classify IRS 29N as an early type star according to \citet{pau06} 
although \cite{fig03} classifies it as a late type star.}

\tablenotetext{a}{The magnitudes are $Q_{j}$ corrected 
using 7 bright non-variable stars, and
the uncertainties do not include the 5 \% absolute calibration uncertainties. 
Comparison to other sources requires adding them in quadrature.
Uncertainties are calculated as the standard deviation of the mean.}
\tablenotetext{b}{Spectroscopic identification by \citet{eis05}. }
\tablenotetext{c}{Spectroscopic identification by \citet{pau06}. }
\tablenotetext{d}{Spectroscopic identification by \citet{fig03}. }
\tablenotetext{e}{Spectroscopic identification by \citet{ott03}.
We denote sources with clear CO or He lines as Early and Late respectively.}
\tablenotetext{f}{Identification based on the interpretation by \citet{gen03} of $m(CO)$ index of \citet{ott03} where
\citet{gen03} identify stars with $m(CO) \geq 0.04$ as late type stars and stars with $m(CO) < 0.04$ 
as early type stars.}
\tablenotetext{g}{Identified as nonvariable by \citet{ott99}. }
\tablenotetext{h}{Identified as possibly variable by \citet{ott99}. }
\tablenotetext{i}{IRS 16CC appears to be variable in the L-band as discussed in \S3.1. }

\end{deluxetable*}

\clearpage
\LongTables
\begin{deluxetable*}{lrccrrcr}
\tabletypesize{\scriptsize}
\tablecaption{List of Non-variable Stars
\label{tab3}}
\tablewidth{0pt}
\tablehead{
\colhead{Star ID} & 
\colhead{Other ID} & 
\colhead{K}   & 
\colhead{p}   & 
\colhead{$\Delta$R.A.} &
\colhead{$\Delta$Dec.} &
\colhead{Nights} &
\colhead{Type} \\
\colhead{} &
\colhead{} &
\colhead{(mag)} &
\colhead{(arcsec)} &
\colhead{(arcsec)} &
\colhead{(arcsec)} &
\colhead{(days)} &
\colhead{}
}

\startdata
IRS16NE &E39 &9.00$ \pm $0.05 &3.06 &2.85 &1.10 &35 &Ofpe/WN9\tablenotemark{c}\tablenotemark{g} \\
IRS16C &E20 &9.83$ \pm $0.05 &1.23 &1.13 &0.50 &50 &Ofpe/WN9\tablenotemark{c}\tablenotemark{g}\tablenotemark{i} \\
S2-17 &E34:GEN+1.27-1.87 &10.90$ \pm $0.07 &2.23 &1.27 &-1.84 &35 &B0.5-1 I\tablenotemark{c}\tablenotemark{g} \\
IRS16SW-E &E32:16SE1 &11.00$ \pm $0.08 &2.15 &1.85 &-1.11 &50 &WC8/9\tablenotemark{c}\tablenotemark{g} \\
IRS29S &\nodata &11.31$ \pm $0.06 &2.08 &-1.86 &0.93 &30 &K3 III\tablenotemark{d} \\
S1-24 &E26:GEN+0.76-1.55 &11.64$ \pm $0.07 &1.72 &0.73 &-1.55 &45 &O8-9.5 I\tablenotemark{c}\tablenotemark{g} \\
S2-16 &E35:29NE1 &11.85$ \pm $0.08 &2.29 &-1.01 &2.05 &33 &WC8/9\tablenotemark{c}\tablenotemark{g} \\
S1-23 &GEN-0.90-1.46 &11.86$ \pm $0.09 &1.73 &-0.92 &-1.46 &33 &Late\tablenotemark{e}\tablenotemark{g} \\
S3-2 &GEN+3.07+0.56 &12.00$ \pm $0.11 &3.09 &3.03 &0.60 &31 &Early\tablenotemark{f} \\
S2-6 &E30:GEN+1.60-1.36 &12.06$ \pm $0.08 &2.07 &1.59 &-1.31 &50 &O8.5-9.5 I\tablenotemark{c} \\
S1-3 &E15:GEN+0.57+0.84 &12.10$ \pm $0.06 &0.99 &0.46 &0.88 &50 &Early\tablenotemark{f}\tablenotemark{h} \\
S3-5 &E40:16SE &12.15$ \pm $0.09 &3.16 &2.95 &-1.13 &21 &WN5/6\tablenotemark{c}\tablenotemark{g} \\
S2-8 &W2 &12.24$ \pm $0.08 &2.16 &-1.99 &0.84 &23 &Early\tablenotemark{f} \\
S1-17 &GEN+0.55-1.45 &12.51$ \pm $0.08 &1.52 &0.50 &-1.44 &49 &Late\tablenotemark{e}\tablenotemark{g} \\
S1-4 &GEN+0.77-0.71 &12.53$ \pm $0.07 &1.02 &0.77 &-0.66 &50 &Early\tablenotemark{f} \\
S2-19 &E36:GEN+0.53+2.27 &12.62$ \pm $0.11 &2.34 &0.42 &2.30 &33 &O9-B0 I?\tablenotemark{c}\tablenotemark{g} \\
S1-20 &GEN+0.41+1.59 &12.70$ \pm $0.11 &1.66 &0.37 &1.61 &49 &Late\tablenotemark{e}\tablenotemark{g} \\
S1-22 &E25:W14 &12.72$ \pm $0.08 &1.72 &-1.65 &-0.51 &42 &O8.5-9.5 I?\tablenotemark{c}\tablenotemark{g} \\
S1-5 &GEN+0.43-0.96 &12.78$ \pm $0.04 &0.98 &0.37 &-0.91 &50 &Late\tablenotemark{e}\tablenotemark{g} \\
S1-14 &E22:W10 &12.82$ \pm $0.07 &1.40 &-1.37 &-0.30 &46 &O8-9.5 III/I\tablenotemark{c}\tablenotemark{g} \\
S3-6 &GEN+3.26+0.08 &12.82$ \pm $0.05 &3.22 &3.22 &0.09 &17 &Late\tablenotemark{e}\tablenotemark{g} \\
S2-22 &GEN+2.37-0.29 &12.92$ \pm $0.04 &2.33 &2.32 &-0.22 &50 &Late\tablenotemark{e}\tablenotemark{g} \\
S2-38 &\nodata &12.93$ \pm $0.09 &2.12 &2.04 &0.58 &44 &Late\tablenotemark{f} \\
S2-31 &GEN+2.91-0.20 &13.06$ \pm $0.10 &2.84 &2.83 &-0.15 &42 &Late\tablenotemark{e}\tablenotemark{g} \\
S1-34 &\nodata &13.20$ \pm $0.14 &1.29 &0.86 &-0.96 &50 &\nodata \\
S2-5 &GEN+1.91-0.86 &13.32$ \pm $0.05 &2.05 &1.89 &-0.80 &50 &Early\tablenotemark{f} \\
S1-68 &\nodata &13.38$ \pm $0.06 &1.97 &1.89 &-0.55 &50 &\nodata \\
S2-21 &GEN-1.70-1.65 &13.47$ \pm $0.10 &2.36 &-1.70 &-1.65 &12 &Early\tablenotemark{f} \\
S0-13 &GEN+0.59-0.47 &13.49$ \pm $0.04 &0.69 &0.54 &-0.42 &50 &Late\tablenotemark{e}\tablenotemark{g} \\
S1-25 &GEN+1.69-0.66 &13.54$ \pm $0.05 &1.76 &1.65 &-0.60 &50 &Late\tablenotemark{e} \\
S2-26 &\nodata &13.60$ \pm $0.13 &2.56 &0.69 &2.47 &20 &Late\tablenotemark{e} \\
S0-15 &E16:W5 &13.70$ \pm $0.07 &0.98 &-0.93 &0.29 &48 &O9-9.5 V\tablenotemark{c} \\
S0-14 &E14:W9 &13.72$ \pm $0.08 &0.83 &-0.78 &-0.27 &50 &O9.5-B2 V\tablenotemark{c} \\
S1-19 &GEN+0.38-1.58 &13.82$ \pm $0.12 &1.62 &0.36 &-1.58 &46 &Early\tablenotemark{f} \\
S2-2 &GEN-0.54+2.00 &14.07$ \pm $0.11 &2.12 &-0.59 &2.03 &41 &Late\tablenotemark{f} \\
S0-2 &E1:S2 &14.16$ \pm $0.08 &0.12 &-0.07 &0.10 &33 &B0-2 V\tablenotemark{b} \\
S1-8 &E18:W11 &14.19$ \pm $0.11 &1.08 &-0.67 &-0.85 &49 &OB\tablenotemark{c} \\
S1-15 &W4 &14.21$ \pm $0.10 &1.46 &-1.37 &0.52 &47 &Late\tablenotemark{f} \\
S0-6 &S10 &14.26$ \pm $0.09 &0.39 &0.07 &-0.38 &49 &Late\tablenotemark{f} \\
S1-49 &\nodata &14.26$ \pm $0.13 &1.66 &-1.65 &0.15 &23 &\nodata \\
S1-13 &W12 &14.27$ \pm $0.12 &1.42 &-1.10 &-0.90 &46 &Early\tablenotemark{f} \\
S2-47 &\nodata &14.29$ \pm $0.08 &2.26 &2.20 &-0.49 &48 &Early\tablenotemark{f} \\
S0-9 &S11 &14.31$ \pm $0.08 &0.55 &0.14 &-0.53 &49 &Early\tablenotemark{f} \\
S0-12 &W6 &14.38$ \pm $0.06 &0.68 &-0.57 &0.37 &49 &Late\tablenotemark{f} \\
S2-3 &W15 &14.48$ \pm $0.09 &2.09 &-1.54 &-1.41 &23 &Late\tablenotemark{f} \\
S0-4 &E10:S8 &14.49$ \pm $0.11 &0.37 &0.32 &-0.19 &49 &B0-2 V\tablenotemark{b} \\
S0-3 &E6:S4 &14.50$ \pm $0.14 &0.25 &0.22 &0.13 &19 &B0-2 V\tablenotemark{b} \\
S2-75 &\nodata &14.52$ \pm $0.12 &2.78 &2.65 &-0.85 &40 &\nodata \\
S2-69 &\nodata &14.57$ \pm $0.14 &2.64 &-0.91 &2.48 &13 &Early\tablenotemark{f} \\
S3-4 &\nodata &14.61$ \pm $0.20 &3.14 &3.10 &-0.47 &20 &Early\tablenotemark{f} \\
S0-1 &E4:S1 &14.67$ \pm $0.11 &0.14 &-0.11 &-0.09 &49 &B0-2 V\tablenotemark{b} \\
S2-23 &\nodata &14.72$ \pm $0.13 &2.43 &1.64 &1.80 &39 &Late\tablenotemark{f} \\
S1-55 &\nodata &14.80$ \pm $0.39 &1.69 &1.58 &0.59 &41 &\nodata \\
S1-50 &\nodata &14.82$ \pm $0.37 &1.67 &1.51 &0.72 &41 &\nodata \\
S1-52 &\nodata &14.83$ \pm $0.29 &1.66 &-0.02 &1.66 &42 &\nodata \\
S1-10 &W8 &14.88$ \pm $0.12 &1.15 &-1.15 &-0.04 &42 &\nodata \\
S1-2 &E17:GEN-0.06-1.01 &14.90$ \pm $0.12 &1.00 &-0.05 &-1.00 &45 &Early\tablenotemark{f} \\
S1-33 &\nodata &15.01$ \pm $0.10 &1.25 &-1.24 &-0.07 &40 &\nodata \\
S1-58 &\nodata &15.04$ \pm $0.35 &1.77 &-1.48 &0.98 &37 &\nodata \\
S1-51 &\nodata &15.05$ \pm $0.15 &1.66 &-1.65 &-0.20 &39 &Early\tablenotemark{f} \\
S2-86 &\nodata &15.05$ \pm $0.23 &2.99 &2.68 &-1.33 &24 &Early\tablenotemark{f} \\
S3-3 &\nodata &15.06$ \pm $0.33 &3.12 &3.06 &-0.62 &19 &Early\tablenotemark{f} \\
S2-30 &\nodata &15.13$ \pm $0.19 &2.88 &2.88 &0.00 &23 &\nodata \\
S1-18 &\nodata &15.14$ \pm $0.15 &1.66 &-0.73 &1.49 &39 &Early\tablenotemark{f} \\
S0-5 &E9:S9 &15.17$ \pm $0.16 &0.36 &0.18 &-0.31 &43 &B0-2 V\tablenotemark{b} \\
S0-31 &E13 &15.20$ \pm $0.23 &0.66 &0.49 &0.45 &40 &B V\tablenotemark{c} \\
S2-34 &\nodata &15.21$ \pm $0.22 &2.04 &1.79 &0.98 &43 &\nodata \\
S0-26 &E8:S5 &15.27$ \pm $0.22 &0.39 &0.36 &0.16 &40 &B4-9 V\tablenotemark{b} \\
S1-44 &\nodata &15.28$ \pm $0.41 &1.61 &0.26 &1.59 &40 &\nodata \\
S3-16 &\nodata &15.30$ \pm $0.20 &3.15 &3.02 &-0.88 &14 &Late\tablenotemark{f} \\
S2-82 &\nodata &15.30$ \pm $0.29 &2.88 &2.87 &0.08 &27 &Late\tablenotemark{f} \\
S2-12 &\nodata &15.33$ \pm $0.17 &2.07 &1.68 &1.22 &38 &Late\tablenotemark{f} \\
S1-32 &\nodata &15.33$ \pm $0.13 &1.13 &-0.93 &-0.64 &42 &\nodata \\
S1-39 &\nodata &15.35$ \pm $0.16 &1.45 &-0.53 &-1.35 &30 &Early\tablenotemark{f} \\
S0-11 &E12:S7 &15.36$ \pm $0.13 &0.53 &0.53 &-0.01 &40 &B V\tablenotemark{c} \\
S0-18 &S18 &15.36$ \pm $0.13 &0.43 &-0.09 &-0.42 &45 &\nodata \\
S1-35 &\nodata &15.36$ \pm $0.19 &1.27 &-1.24 &-0.25 &38 &\nodata \\
S2-63 &\nodata &15.39$ \pm $0.28 &2.56 &-0.69 &2.47 &16 &Early\tablenotemark{f} \\
S1-54 &\nodata &15.41$ \pm $0.23 &1.68 &-1.53 &0.70 &36 &\nodata \\
S1-62 &\nodata &15.41$ \pm $0.31 &1.82 &0.51 &1.74 &37 &\nodata \\
S1-53 &\nodata &15.44$ \pm $0.15 &1.68 &1.68 &-0.09 &30 &\nodata \\
S2-61 &\nodata &15.46$ \pm $0.17 &2.54 &2.46 &-0.64 &38 &\nodata \\
S2-46 &\nodata &15.47$ \pm $0.30 &2.18 &2.08 &-0.64 &38 &\nodata \\
S2-73 &\nodata &15.50$ \pm $0.32 &2.72 &2.21 &-1.58 &29 &\nodata \\
S1-6 &\nodata &15.54$ \pm $0.13 &1.15 &-0.91 &0.71 &34 &Early\tablenotemark{f} \\
S1-48 &\nodata &15.54$ \pm $0.23 &1.62 &-0.60 &-1.50 &22 &\nodata \\
S0-7 &E11:S6 &15.55$ \pm $0.27 &0.45 &0.44 &0.11 &35 &B V\tablenotemark{c} \\
S0-19 &E5 &15.56$ \pm $0.19 &0.19 &-0.09 &0.17 &28 &B4-9 V\tablenotemark{b} \\
S1-64 &\nodata &15.57$ \pm $0.36 &1.91 &0.60 &1.81 &34 &\nodata \\
S2-80 &\nodata &15.57$ \pm $0.27 &2.86 &2.22 &1.80 &20 &Early\tablenotemark{f} \\
S2-40 &\nodata &15.58$ \pm $0.18 &2.15 &1.68 &1.34 &35 &Early\tablenotemark{f} \\
S2-42 &\nodata &15.61$ \pm $0.28 &2.11 &0.41 &2.07 &26 &\nodata \\
S1-27 &\nodata &15.62$ \pm $0.27 &1.09 &-1.07 &0.24 &39 &\nodata \\
S1-26 &\nodata &15.62$ \pm $0.13 &1.02 &-0.95 &0.38 &40 &\nodata \\
S1-37 &\nodata &15.63$ \pm $0.16 &1.42 &-1.34 &0.47 &36 &\nodata \\
S1-47 &\nodata &15.67$ \pm $0.23 &1.63 &-1.57 &0.45 &34 &\nodata \\
S0-16 &E2 &15.68$ \pm $0.19 &0.08 &0.05 &0.06 &18 &B4-9 V\tablenotemark{b} \\
S1-59 &\nodata &15.70$ \pm $0.25 &1.86 &0.01 &1.86 &27 &\nodata \\
S1-31 &GEN-0.91+0.44 &15.70$ \pm $0.18 &1.14 &-0.99 &0.57 &31 &\nodata \\
S0-29 &\nodata &15.72$ \pm $0.50 &0.54 &0.25 &-0.48 &21 &\nodata \\
S2-83 &\nodata &15.74$ \pm $0.19 &2.94 &2.87 &-0.63 &16 &\nodata \\
S0-27 &\nodata &15.74$ \pm $0.18 &0.54 &0.13 &0.52 &32 &\nodata \\
S1-36 &\nodata &15.75$ \pm $0.27 &1.36 &-0.67 &-1.18 &30 &\nodata \\
S2-37 &\nodata &15.77$ \pm $0.31 &2.11 &0.04 &2.11 &29 &\nodata \\
S0-8 &E7 &15.79$ \pm $0.14 &0.47 &-0.32 &0.34 &29 &B4-9 V\tablenotemark{b} \\
S1-7 &\nodata &15.81$ \pm $0.14 &1.12 &-1.00 &-0.50 &34 &\nodata \\
S1-65 &\nodata &15.82$ \pm $0.15 &1.93 &1.43 &1.29 &31 &Early\tablenotemark{f} \\
S0-28 &S19 &15.85$ \pm $0.18 &0.61 &-0.18 &-0.58 &30 &\nodata \\
S2-64 &\nodata &15.85$ \pm $0.59 &2.56 &2.54 &0.31 &16 &\nodata \\
S0-36 &\nodata &15.85$ \pm $0.28 &1.03 &-0.60 &-0.84 &29 &\nodata \\
S0-20 &E3 &15.86$ \pm $0.20 &0.21 &-0.18 &-0.10 &33 &B4-9 V\tablenotemark{b} \\
S1-40 &\nodata &15.95$ \pm $0.23 &1.50 &-1.36 &-0.64 &26 &\nodata \\
S1-61 &\nodata &15.99$ \pm $0.38 &1.76 &-1.43 &-1.03 &21 &\nodata \\
S2-52 &\nodata &16.02$ \pm $0.20 &2.37 &2.37 &-0.07 &21 &\nodata \\
S1-42 &\nodata &16.13$ \pm $0.22 &1.60 &0.94 &1.29 &19 &\nodata \\

\enddata

\tablecomments{See notes from table \ref{tab1}.
}
\tablenotetext{a}{The magnitudes are $Q_{j}$ corrected 
using 7 bright non-variable stars, and
the uncertainties do not include the 5 \% calibration uncertainties. 
Comparison to other sources requires adding them in quadrature.}
\tablenotetext{b-h}{Type from \citet{pau06}, \citet{fig03}, \citet{eis05}, and \citet{ott03} 
as described in table \ref{tab1}.}
\tablenotetext{g}{Identified as nonvariable by \citet{ott99}. }
\tablenotetext{h}{Identified as possibly variable by \citet{ott99}. }
\tablenotetext{i}{This is our main calibration star and is included in this table only for completeness.}
\tablenotetext{j}{The 7 least variable bright stars detected in all images used
for scaling of the photometry in order to reduce
the fluctuations induced by measurement errors on IRS 16C (see \S 3.1).}

\end{deluxetable*}

\clearpage

\noindent  aperture photometry), 
the time baseline (10 years vs. 5 years), and the angular
resolution (0$\farcs$05  vs. 0$\farcs$13 ). 
Since our study covers twice the time baseline, we can pick out variations 
on longer time scales, which helps explain our additional variables. 
Also, the high stellar crowding makes the area we 
observe the most uncertain region for the lower resolution \citeauthor{ott99} study.
Only one star, S1-3, in our non-variable sample is identified as variable by \citeauthor{ott99},
and it is very close to their threshold for variability.

\begin{figure}
\plotone{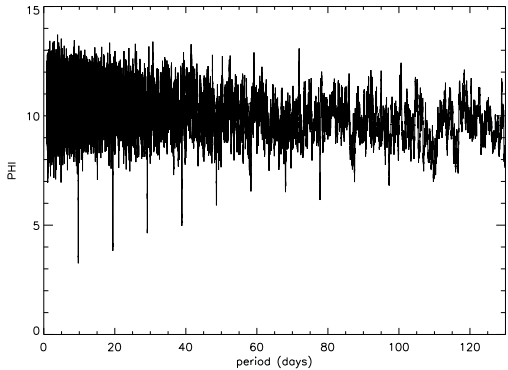}
\caption{Determination of IRS 16SW's period using Dworetsky's string length method \cite{dwor83}. 
Phi is the string length where shorter lengths and therefore smaller values of Phi indicate a periodic
signal. Note that the other harmonics are also found at multiples of the original period.
 \label{dwor16sw}}
\end{figure}

\begin{figure*}
\epsscale{.9}
\plotone{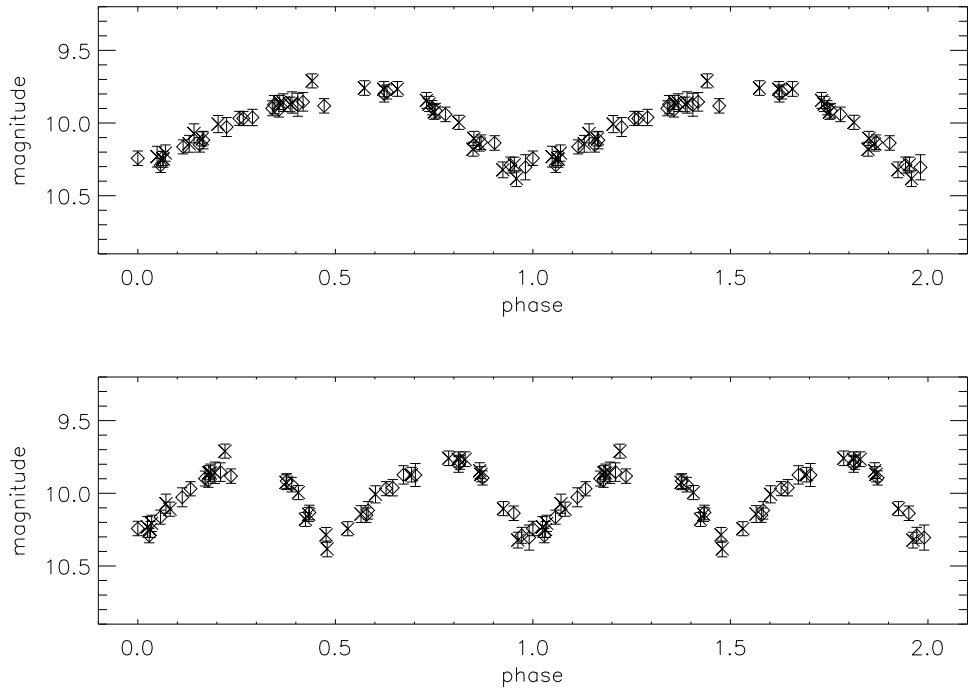}
\caption{The phased light curve of IRS 16SW at its period of 9.724 (top panel) and 19.448 days
(bottom panel). The points are plotted twice for clarity. Both phased diagrams
show a rise-time that is $\sim1.6$ times longer than the fall-time. 
The similarity of the first half of the data set (1995-2000; diamonds) to the second half
(2001-2005; crosses) shows that the asymmetry has been constant over the past ten years. 
This effect could be produced by tidal deformation of two equal mass 
stars in an eclipsing binary star system. 
\label{phase16sw}}
\epsscale{1.0}
\end{figure*}

\subsubsection{Variability Characterization and Periodicity Search}
We attempt to characterize the minimum time-scale for variation by searching for 
daily and monthly variability using KS tests similar to the test for variability in \S 3.2.1 
where we adopt as our model a non-variable light curve with 
gaussian-distributed uncertainties. For the daily variations, we group the consecutive 
nights in pairs and examine the distribution of the pairs in sets $i$ for each star of
$X_{i} = \frac{Flux_{j}-Flux_{k}}{\sqrt{\sigma_{Flux_{j}}^{2} + \sigma_{Flux_{k}}^{2}}}$
where $Flux_{j}$ and $Flux_{k}$ are the fluxes of stars in 
images of consecutive epochs j and k respectively,
$\sigma_{Flux_{j}}$ and $\sigma_{Flux_{k}}$ are the corresponding uncertainties.
For the monthly variation, all measurements made within days of each other are averaged together 
and the pairs separated by one month are examined with the same KS test.
The only stars showing daily variability in excess of $3\sigma$ is IRS 16SW, and the 
stars showing monthly variations are IRS 16SW and S2-36.

The light curves of the variable stars are searched for periodicities using three different methods.
First, the Lomb and Scargle periodogram technique \citep{lomb76, scargle82, press92},
which fits fourier components to the data points, is applied and is expected to yield a
larger power spectral density at intrinsic harmonics of a 
data set in which there is a periodic signal. Second, 
Dworetsky's \citeyearpar{dwor83} string length method, 
a variant of the Lafler-Kinman method \citeyearpar{laf65},
phases the data for every possible period and then sums over the 
total separation between points in phase space, with the best period and its aliases 
corresponding to the smallest lengths. 
Third, Stetson's \citeyearpar{stet96} string length technique
is similar to Dworetsky's, but also weights these lengths by their
uncertainties and how close in phase the points are. 
Our criterion for considering a star periodic is that it show similar periods from 
all three techniques.
The periodicity search shows only one periodic star: IRS 16SW. We find 
a photometric period of $9.724 \pm 0.001$ days, which is consistent with \cite{ott99}, \cite{depoy04}
and the reanalysis of the \citeauthor{ott99} data in \citet{mar06}.
Figure \ref{dwor16sw} shows the determination of IRS 16SW's period using 
Dworetsky's (1983) string length algorithm, showing the 9.724 day period and 
its other harmonics at multiples of its period.
The top panel in Figure \ref{phase16sw} shows the phased light curve of IRS 16SW 
at 9.724 days and depicts a clearly periodic signal with an amplitude of $\sim0.55$ mag.
The phased light curve of IRS 16SW is asymmetric (see Figure \ref{phase16sw})
with a rise-time that is $\sim1.6$ times longer than the fall-time.

\begin{figure*}
\epsscale{.9}
\plotone{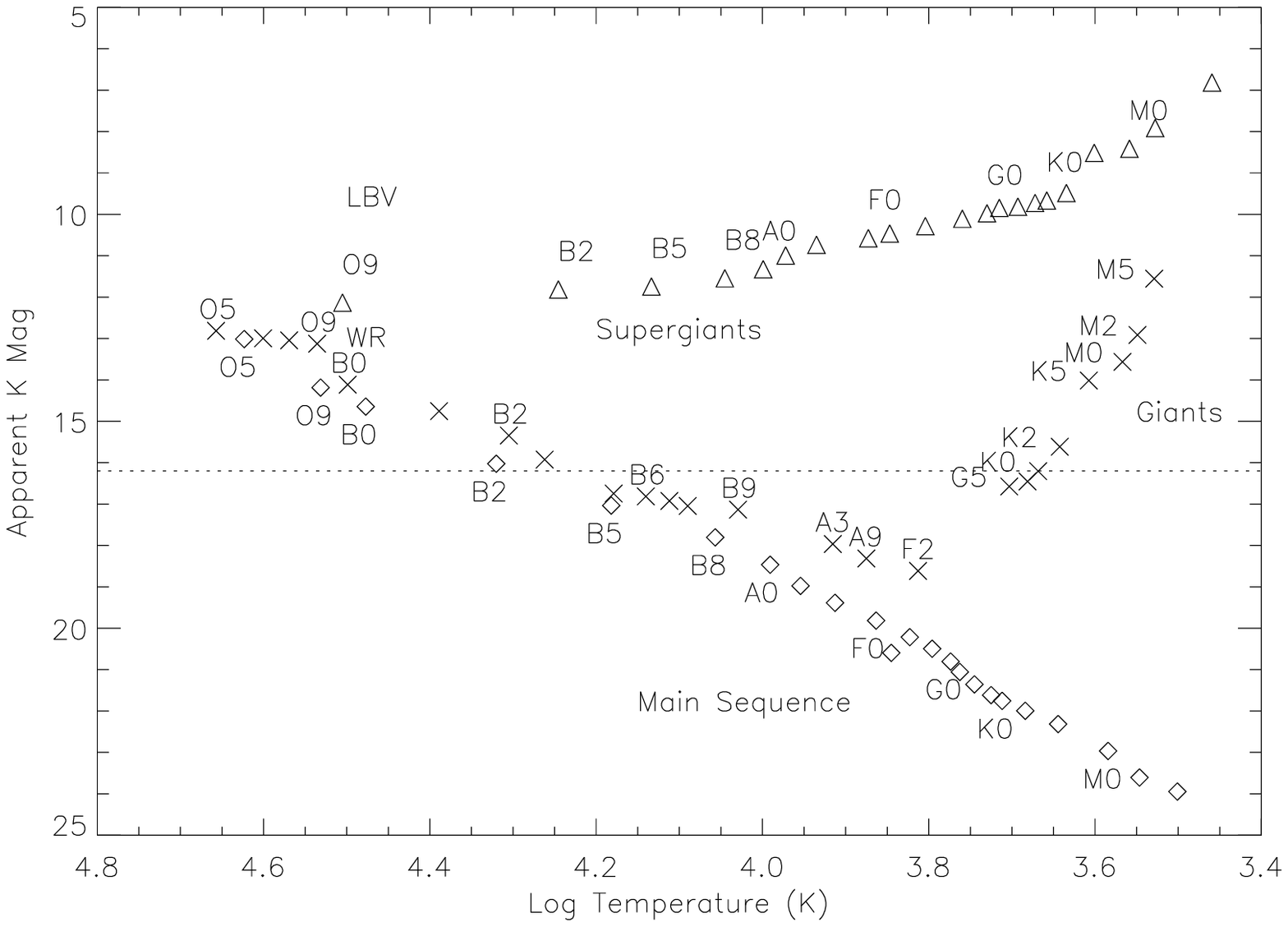}
\caption{Apparent K magnitudes versus spectral type assuming
a distance of $r=8.0$ kpc \citep{reid93} and an extinction $A_{K}$=3.3 mag
\citep{blum96a} based on
compilations of data from \citet{cox00} where available, 
and by \citep{jag87, weg94, weg06, vac96, gren85,  black94, theo91} for Giants of type O-F . 
The lower dotted line is the magnitude limit of this sample, and
the locations of WR and LBV stars are highly variable and average locations
are marked for reference only. The triangles represent supergiant stars, 
the crosses represent giant stars, and 
the diamonds represent main sequence stars.
\label{spec}}
\epsscale{1.0}
\end{figure*}

\section{Discussion}

The 16 variable stars identified in this study 
cover a wide variety of different types of stars, as we only limited
our search by location and brightness.
As Figure \ref{spec} shows, based on the K magnitudes alone, this sample 
is expected to contain early-type (O \& B) main
sequence stars, late-type (K \& M) giant stars, and most types
of supergiants.  
Fortunately, all but two of the variable stars have 
spectral classifications (see column 10 in Table \ref{tab2}).
While 9 of the variable stars are securely identified from 
spectroscopic work \citep{pau06, eis05, fig03, ott03},
an additional 5 stars are classified on the basis of narrowband photometry of CO absorption
\citep{ott03, gen03}. 
In summary, four stars are LBV candidates (IRS 16SW, 16NW, 16NE, \& 16C; see \S4.1)
one is a WC9 (IRS 29N; see \S4.2),
four are OB supergiants (S2-4,  S1-12, S2-7, \& IRS 16CC; see \S4.3),  
one is an O giant (S1-21),  
five more are classified as some sort of early-type star from narrowband filter photometric
measurements (S1-1, S2-36, S0-32, S2-58, S1-45), and one is classified as a late-type
star from narrowband filter photometric measurements (S2-11; see \S4.3).  
The variability of stars in each spectral
classification is discussed in turn below, along with a discussion of the possibility of external agents causing variability.

\subsection{Ofpe/WN9 Stars}

\subsubsection{Luminous Blue Variable Candidates}

\begin{figure*}
\epsscale{0.9}
\plotone{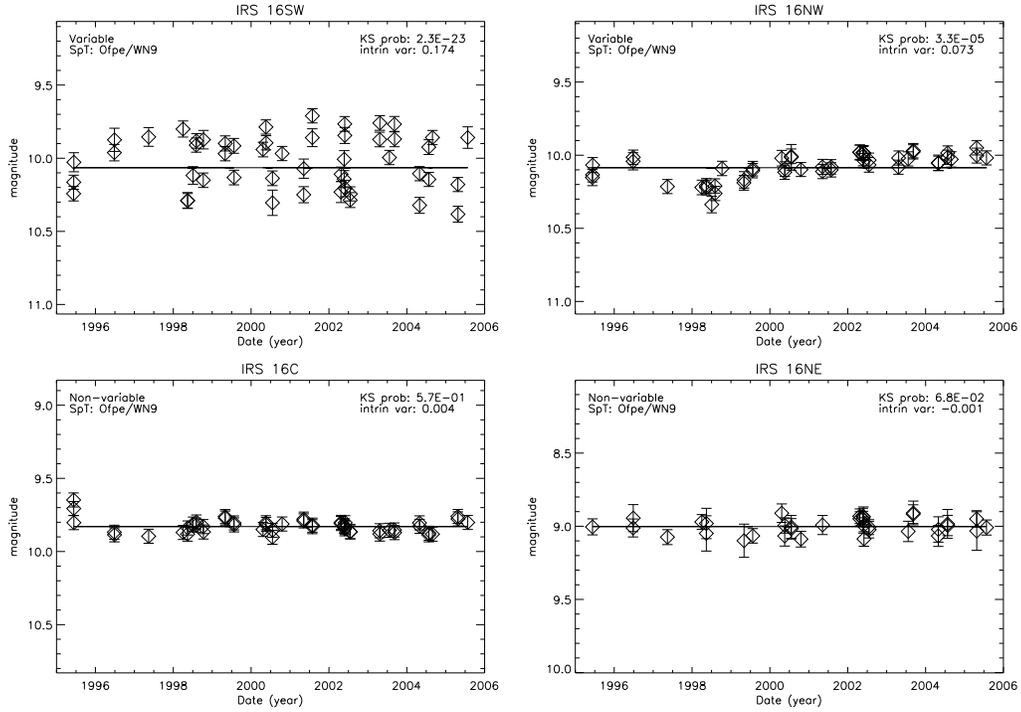}
\caption{Light curves of the four stars in our sample that are classified as candidate LBVs 
(IRS 16SW, 16NW, 16NE, 16C). The top two panels denote the two variable stars, while
the bottom  two panels are non-variable stars. The given intrinsic variance is in magnitudes.
\label{lclbv}}
\epsscale{1.0}
\end{figure*}

Four stars in our sample are Ofpe/WN9 stars and 
have been previously classified as candidate LBVs
(IRS 16NE, 16C, 16SW, and 16NW) based on their bright luminosity, 
their narrow emission lines, and their proximity and similarity
to IRS 34W \citep{clark05, pau04b, trip06}
\footnote{The classification of IRS 34W as an LBV is based on its bright luminosity, 
narrow emission lines, along with a multi-year 
obscuration event \citep{pau04b, trip06}.
However, more recent studies have cast doubt on IRS 34W's categorization as an LBV since 
it lacks spectroscopic variability and since 
the eruption event responsible for the obscuration event
was not observed\citep{trip06}.}.
In our observations, both IRS 16NE and 16C are non-variable over a ten-year 
time frame to within our uncertainties. 
The other two LBV candidates (IRS 16NW and 16SW) show variability, but not the
characteristic LBV eruptions, which in the context of this study are the 
$\Delta M_{v} \simeq 1-2$ mag events occurring every 10-40 years and lasting
as long as several years \citep{hum94}.
IRS 16SW has periodic variability 
that is explained by an eclipsing binary system (see \S 4.1.2 below) and
IRS 16NW has an overall flat light curve with a decrease in brightness 
($\Delta m_{K}  \simeq 0.2$) mag between 1997 and 1999 (see Figure \ref{lclbv}). 
The apparent dimming of IRS 16NW can be explained by ejected 
circumstellar material obscuring the star, 
with an amplitude that is smaller than is characteristic of a typical LBV.
These variations do not require it to be an LBV, just that it has strong stochastic winds.
While none of these stars shows the classic characteristics of LBVs,
our time baseline is too short to rule them out as LBVs as they may be in a quiescent phase. 
Nonetheless, our observations do not demand the complication of multiple
recent star formation events that would be suggested with the presence of both
LBVs and WR stars.

\subsubsection{Asymmetric Periodic Light Variations in IRS 16SW: Tidal Deformation?}

The periodic variation in IRS 16SW has recently been attributed to either
an equal mass contact eclipsing-binary or a massive pulsating star 
\citep{ott99, depoy04, mar06}, although the measurement of a spectroscopic 
radial velocity period by \citet{mar06} strongly suggests it is an eclipsing binary star system.
Also, a re-analysis of the data suggesting
a pulsating star now agrees with IRS 16SW probably being an eclipsing binary \citep{peep06}.
The asymmetries that we observe in the phased light curve of IRS 16SW are difficult 
to explain in the context of an eclipsing binary star system.
Figure \ref{phase16sw} shows the properly phased light curve ($19.448 \pm 0.002$ days),
in which asymmetries in its rise and fall-times are still readily detected. 
The asymmetry is remarkably similar for both halves of the phased light curve. 
While this was not detected in earlier photometric studies\footnote{Our 
light curve is similar to the light curve presented by \citet{ott99}, 
although the asymmetry was not explicitly reported and the reanalysis of the data
by \citep{mar06} does not show the asymmetry.}, our study is likely more sensitive to 
small photometric variations due to our higher angular resolution (see \S 3.2.1).
The observed asymmetry has been sustained over 10 years;
if the asymmetry has been 
changing over time, it would show up as a dispersion 
in the vertical placements of the points in the phase 
diagram that is much larger than what we observe. 
In figure \ref{phase16sw} we explicitly show that the asymmetries are the same
during the first half and second half of the data-set and we see no
period drifts, with the period of the first half and second half 
not differing at the $3\sigma$ confidence level. 
Magnetic hot spots can explain 
light curve asymmetries \citep{dju00, cohen04}, however,
they can not maintain this asymmetry over long periods of time. 
Furthermore, the similarity in the asymmetry between the first and 
second half would require the spots to be the same on both stars. 
Likewise, heating in the contact region of the binary or 
any third light in this region
produces a light curve that is a mirror reflection between the 1st and 
2nd half (see \citealp{mof04}) and is therefore inconsistent with our observations. 
Other heating mechanisms such as irradiation effects are unlikely to be the cause 
of the asymmetry as they barely change the light curve \citep{bau05}. 
Given that this is suspected to be a contact binary, we suggest
that the asymmetric light curve 
may be due to tidal deformations caused
by the proximity of the stars in asynchronous orbits. 
Since these two stars are equal in mass ($\sim50M_{\odot}$),
equal in radius ($66R_{\odot}$) and appear to be in contact \citep{mar06},
they fall within the tidal radius of $\sim163R_{\odot}$, and are likely tidally deformed.
In order to produce the asymmetric light curve, the rotation of the stars
needs to be asynchronous with their orbital periods so that the rotational inertia 
of the stars prevents their tidal bulges from being aligned with the line 
joining the stars' centers of mass.
Two stars in such close contact will eventually synchronize their orbital period with 
their rotation period. 
The synchronization time scale of stars with convective cores and
radiative envelopes is longer than for stars with convective envelopes, albeit
more difficult to calculate \citep{zahn77}. We therefore calculate
the convective envelope synchronization time as a minimum time for synchronization 
based on formalism developed by \citet{zahn77} and find a minimum age of 
$\sim6 \times 10^{8}$ yrs. This time scale is far larger than the lifetime of 
stars as massive as IRS 16SW and asynchronous orbits are therefore acceptable. 
The equality of the two halves of the light curve implies that the rotation rate of the 
two stars is very similar.  Since these two stars are identical in every other respect,
this equality is not surprising. 
It is therefore possible that the asymmetric light curve is due to tidal deformations.

\subsubsection{Eclipsing Binary Fraction}

Stars close to the Galactic center are expected
to have a higher fraction of ellipsoidal and eclipsing variable binaries than 
the stars in the solar neighborhood. 
However, IRS 16SW is the only eclipsing binary star detected in our sample. 
The stars' orbits may be smaller due to 
hardening by encounters with other stars producing tightly bound binaries, 
such that eclipses would be more likely. In addition, collisions and tidal capture produce
binaries, making ellipsoidal variations or eclipses
more probable \citep{tam96}.
Of the 164 Galactic O stars in clusters or associations in the sample by \cite{mason98}, 
50 are confirmed as spectroscopic binaries, 40 are unconfirmed spectroscopic 
binaries, 4 are confirmed eclipsing variables, and 14 are either ellipsoidal 
or eclipsing binaries. This yields local rates of eclipsing O star binaries between 2\% and 11\%.
Our sample contains 11 spectroscopically confirmed O stars, 4 Ofpe/WN9 stars,
and possibly more unconfirmed. If the Galactic center
fraction of eclipsing binaries is similar to the cluster results, we would 
expect on the order of one eclipsing binary. 
Our detection of one 
eclipsing variable star suggests
that the frequency of eclipsing binaries is not
significantly increased at the Galactic center over the local neighborhood. 

\begin{figure*}
\epsscale{0.9}
\plotone{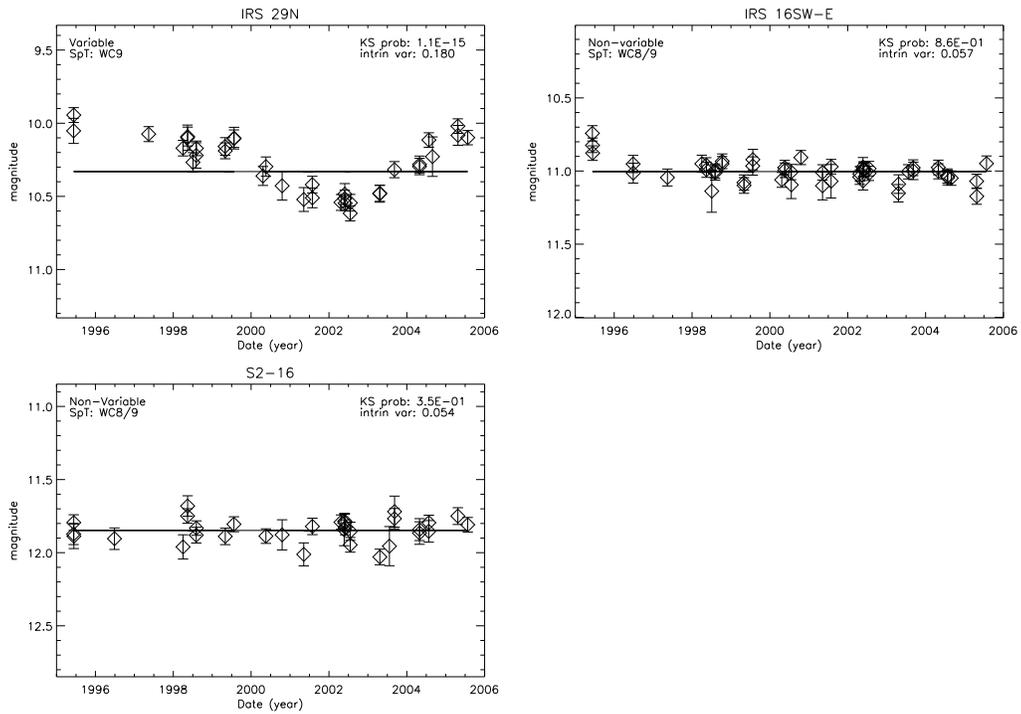}
\caption{Light curves of Wolf-Rayet stars of type WC in our sample. The first panel is the
variable star IRS 29N, which is probably a wind-colliding binary, 
while the other two are non-variable stars. 
The given intrinsic variance is expressed in magnitudes.
\label{lcwc}}
\epsscale{1.0}
\end{figure*}

\subsection{Late-type WC Wolf-Rayet Stars: Variations Associated
with a Wind Colliding Binary}

Three stars in our sample are spectroscopically identified as WC stars 
(IRS 29N, IRS 16SW-E and S2-16) \citep{pau06},
and all three are dust producers as evidenced by their red colors
(K-L $\sim3$; \citealt{blum96a}, \citealt{wrt06}). 
In this study, only IRS 29N is variable (see Fig \ref{lcwc}). 
IRS 29N's intensity shows a gradual drop and then rise in 
brightness of $\Delta m_{K}  \simeq 0.7$ mag over a time scale of 
$\approx 5$ years. Its light curve is similar to the variations seen
in WC stars elsewhere in the Galaxy (e.g. compilation by \citealt{hucht01b}); 
these sources are thought to 
be variable due to periodic or episodic dust production in the wind collision
zone of long period eccentric binary star systems (1000d $<$ P $<$ 10000d) 
during periastron passage. 
When the dust forms, the star exhibits a rising infrared flux followed by fading emission when 
dust formation stops and dust grains are dispersed by stellar winds 
\citep{mof87, white95, veen98, williams92, williams00}. 
Currently, only seven WC stars have been observed to produce dust episodically,
all of which are confirmed or suspected massive binaries with elliptical orbits \citep{hucht01a, williams05, lef05}.
Two of these are known to exhibit pinwheel nebulae, a tell-tale sign of wind-colliding binaries
\citep{tut99, mon99}.
The time-scales and magnitude of the photometric variability of IRS 29N is consistent with it being a wind-colliding binary;
we therefore conclude that IRS 29N is likely to be a wind-colliding binary.

\begin{figure*}
\epsscale{0.9}
\plotone{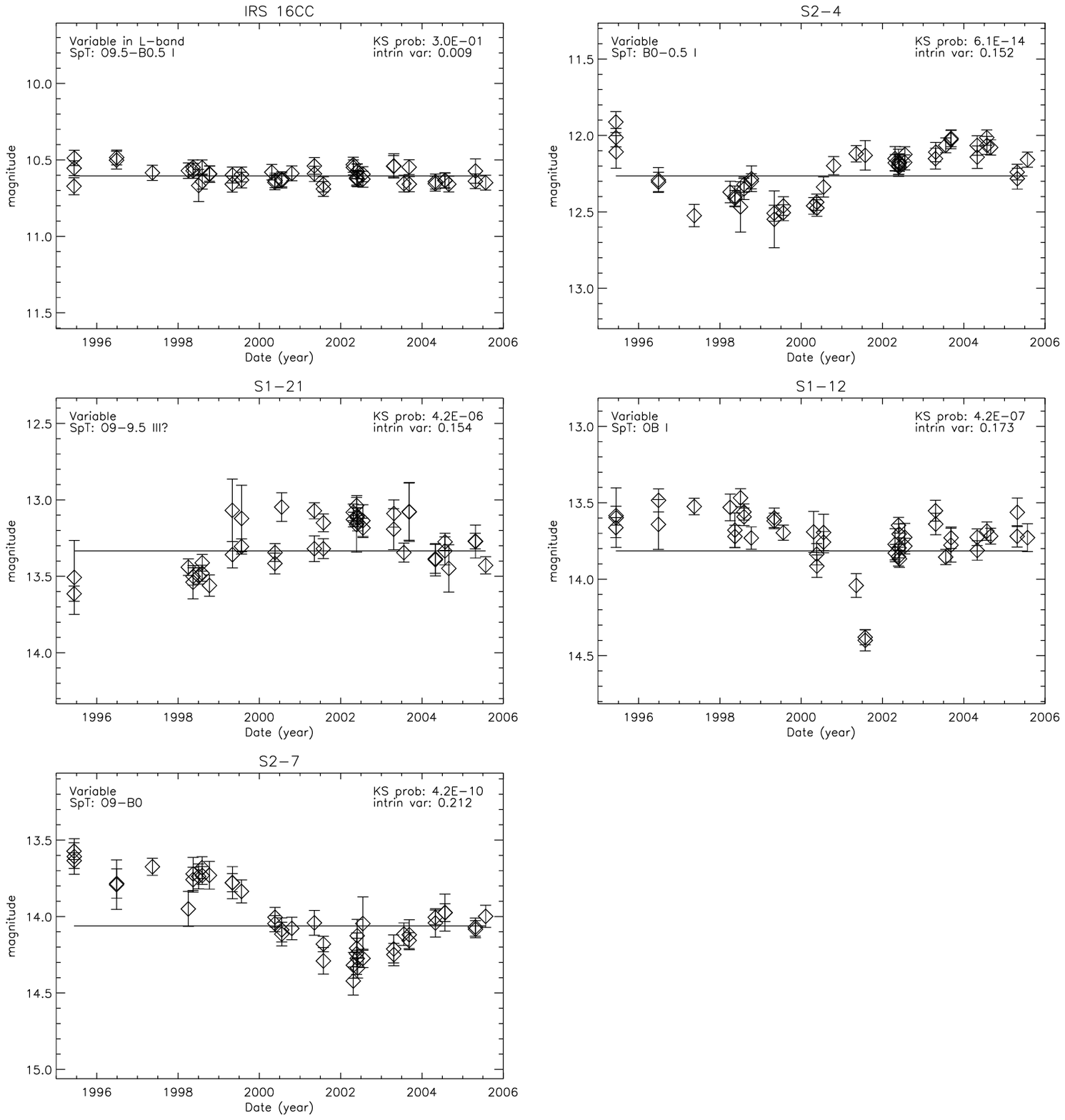}
\caption{Light curves of variable OB stars in our sample. IRS 16CC is not variable in K band, but
shows variability at L (see \S 3.1). We propose that S2-4 and S1-12 are variable due to ejection
of circumstellar material obscuring the stars, while S2-7 is probably a Be star with the formation of
an equatorial disk obscuring the star. The given intrinsic variance is expressed in magnitudes.
\label{lcob}}
\epsscale{1.0}
\end{figure*}

\begin{figure*}
\epsscale{0.9}
\plotone{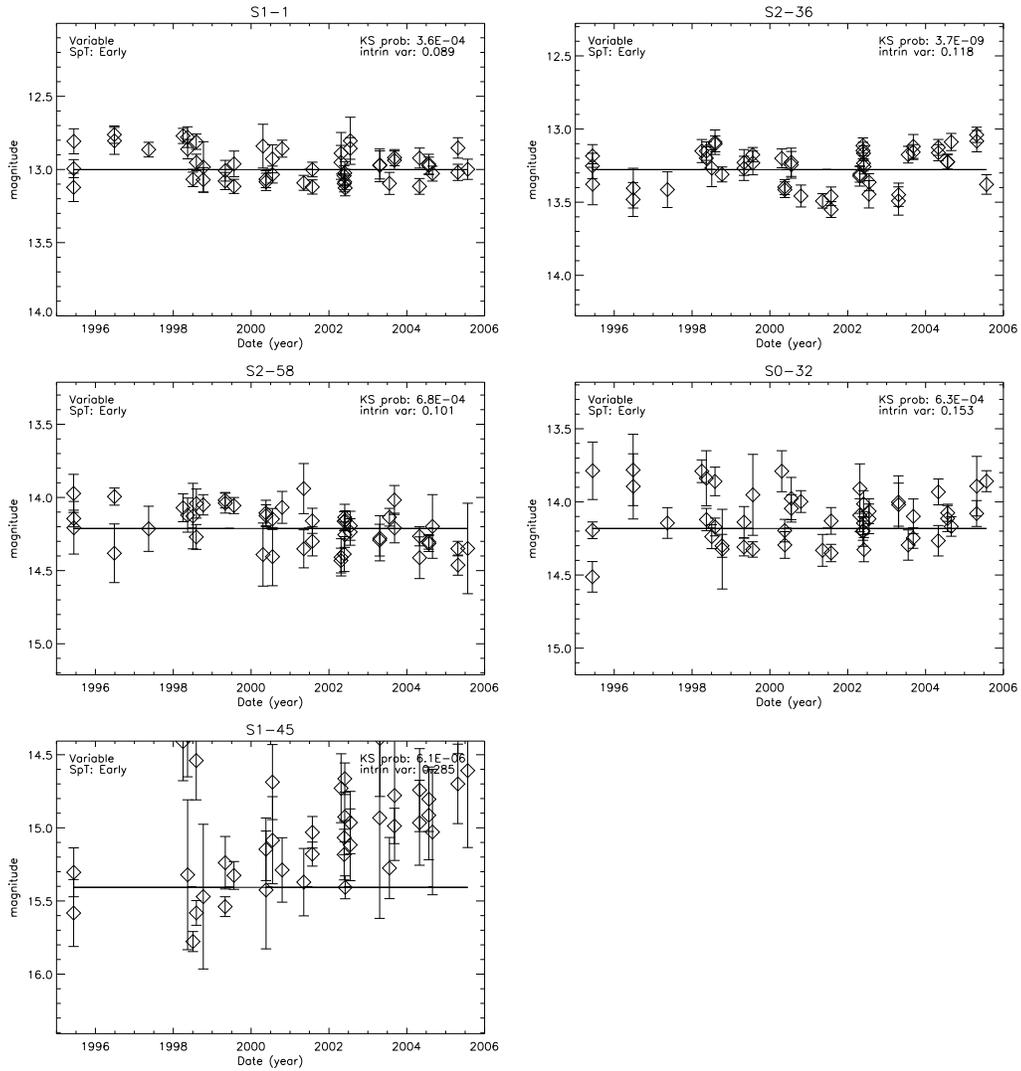}
\caption{Light curves of early-type variable stars in our sample. S2-36 is variable on a 
monthly basis and its variation is therefore probably intrinsic. The variations in S1-45 and S2-58
may be due to obscuration by high density streamers. The given intrinsic variance is expressed
in magnitudes.
\label{lcearly}}
\epsscale{1.0}
\end{figure*}

\clearpage

\subsection{Comments on Other Stars}

Our sample includes at least 10 other variable young massive stars in addition to 
those discussed in \S4.1 \& \S4.2. Five are spectroscopically identified OB stars
(IRS 16CC\footnote{IRS 16CC is not identified as variable in this survey, 
although reported differences in 
L-band magnitudes from previous studies suggest it is variable 
\citep{depoy91, sim96, blum96a, wrt06} (see \S 3.1).}, S2-4, S1-21, 
S1-12, S2-7; see Figure \ref{lcob}) \citep{pau06}, and
five are early-type stars classified as on the basis of narrowband filter photometric measurements 
(S1-1, S2-36, S0-32, S2-58, S1-45; see Figure \ref{lcearly}) \citep{gen03, ott03}.
The majority of these variables are 
likely associated with mass loss, although interstellar extinction could also play a role,
as discussed below. In particular, the two K-band variable OB supergiants each show a dip
(0.3 - 0.9 mag) in their brightness that lasts for 1-6 years. OB supergiants have 
stochastic winds with high mass loss 
rates ($0.2-20\times10^{-6}M_{\odot}$/yr) \citep{mas03};
it is therefore likely the large variations seen are due to ejection of circumstellar material. 
Another example of an OB star with variations potentially due to mass loss is S2-7, which 
is either luminosity class III or V based on its assumed distance. S2-7 shows 
a decrease in luminosity between 2000 and 2005
with $\Delta m_{K}  \simeq 0.5-1.0$ mag. 
This is reminiscent of Be stars, which sometimes
show fading events due to the formation of an equatorial disk on 
time scales of several years \citep{men94, pav97, per01}. 
The remaining young variable stars in our sample have variations that are more
difficult to characterize, although S2-36 seems to have significant variations on monthly 
time-scales (see \S 3.2.2). Nonetheless, it is likely that mass loss also plays a central role in 
generating the observed variations.

\subsection{Interstellar Material}

\subsubsection{Apparent Variations Caused by Stellar Motion Through the Line of Sight Extinction}

Periods of reduced luminosity in stars can also be due to external effects
such as obscuration by foreground interstellar matter. 
The central parsec has several gas patches that are a few arcseconds wide and
that can cause local extinction enhancements of $\sim1$ magnitude in K-band \citep{pau04}. 
As our FOV is only 5$\tt'' \times$5$\tt''$, these patches would cover a sizable region
and many neighboring stars would show similar variations. 
Interstellar material closer to Earth in the line of sight 
is unlikely, as its proximity would make the clouds even larger in 
projection \citep{trip06}. Since neighboring stars do not experience similar 
effects of obscuration, the large scale structure in the interstellar medium
(ISM) is unlikely to be the cause. 

\begin{figure*}
\epsscale{0.9}
\plotone{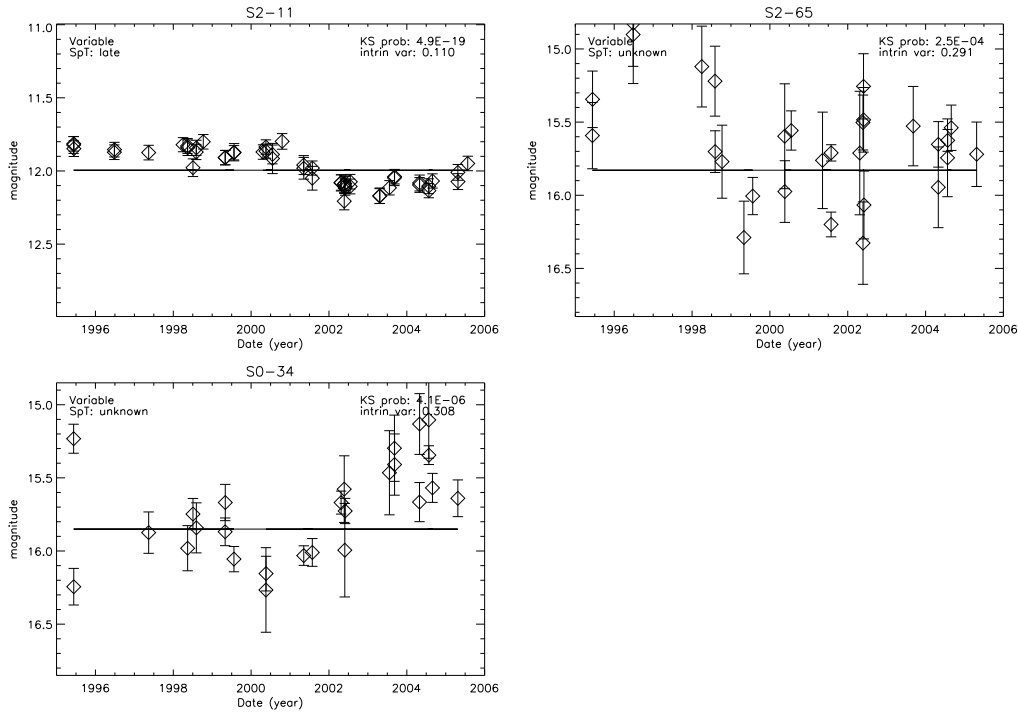}
\caption{Light curves of a late-type star, and two unknown type stars in our sample. 
The late-type star, S2-11, is probably an AGB star of spectral type M3-5 III, while we 
are unable to classify the two stars of unknown spectral type. 
The given intrinsic variance is expressed in magnitudes.
\label{lcmisc}}
\epsscale{1.0}
\end{figure*}

\begin{figure}
\epsscale{1.2}
\plotone{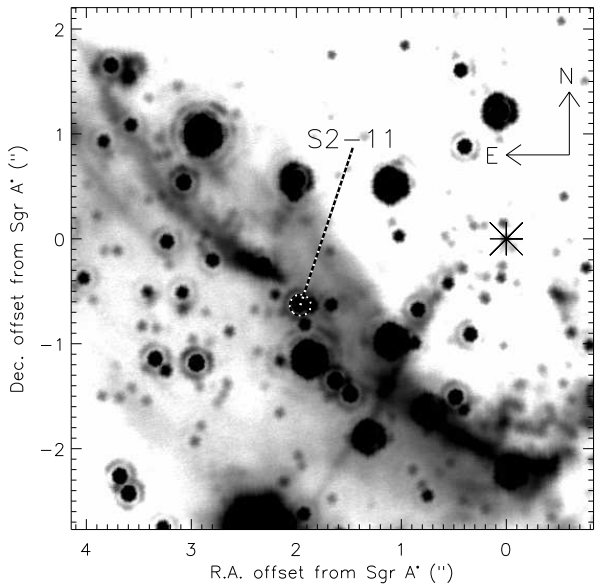}
\caption{
Identification of the source S2-11 from this study overlaid on 
a 5$\tt''$ x 5$\tt''$ region of an LGS L$\tt'$-band ($\lambda_{o}$ = 3.8\micron)
image from \citet{ghez06} taken on June 30, 2005.
The star S2-11 is situated in the middle of the Northern Arm and it is 
therefore likely that the interval of reduced luminosity is due to the stellar motion
through a high density streamer in the line of sight.
The location of Sgr A$^{*}$ is marked with an asterisk.
\label{arm}}
\epsscale{1.0}
\end{figure}

In some cases it is possible that
the variable obscuration is caused by 
the relative motion of the foreground high density streamers 
and the background stars, such as are observed in the L-band
associated with the Northern Arm \citep{clen04, ghez05b, muz06}.
In this area we see unresolved streamers that
are small in projected width ($< 80$ mas). 
These streamers may be due to shocks heating the neighboring dust  
with the streamers tracing thin shells of compressed gas from 
one or several shocks \citep{clen04}. 
These streamers may cause dips in our light curves 
due to small-scale structure obscuration in the line of sight. 
This causes photometric variability due to the relative lateral motion between 
the absorbing feature and the star.
The projected implied width of the small-scale structure is
approximately $\sim10-40$ mas assuming a projected stellar velocity of $\sim5$ mas/yr which is
typical of stars at projected distances of $\sim2\tt''$ from Sgr A$^{*}$.
The three stars whose variability is most likely ascribable to these thin high density streamers are
S2-11, S1-45, S2-58.
The most clear case is the
late-type star S2-11 \citep{ott03},  which shows an interval 
of reduced luminosity between 2001 and 2005 with $\Delta m_{K} \simeq 0.3$ mag 
and is otherwise constant over our time frame (see Figure \ref{lcmisc}).
Using a Galactic center distance of $r=8.0$ kpc \citep{reid93} 
and an extinction $A_{K}$=3.3 mag \citep{blum96a} , we 
determine its spectral type based on luminosity and late-type classification as M3-5 III.
Stars with spectral type M5 and luminosity class III are generally classified as asymptotic giant
branch stars (AGB), but the variations observed are not typical of AGB stars which
have periods between 0.5 - 1.5 yrs \citep{hab96}.
This star is likely obscured by dust given its red color (K-L $\sim2$; \citealt{wrt06}).
It is located 
in the middle of the Northern Arm (see Figure \ref{arm}) and 
its reduced luminosity is likely due to
a high density streamer in the line of sight. 
The two other stars whose variability can probably be attributed to
high density streamers have long term variations over ten years; S1-45
appears to brighten by $\Delta m_{K}  \simeq 0.5-1.0$ mag 
and S2-58 appears to dim by $\Delta m_{K}  \simeq 0.2$ mag (see Figure \ref{lcearly}).
It is also possible that the dips in the light curves
of stars speculated to be variable due to high stellar winds 
such as S2-4, S1-12, S2-7, and S2-36 are actually variable due to obscuration in the line of sight. 
Measurements at multiple wavelengths throughout future variations
would help to establish the role of variable extinction
in the observed K-band variations. Furthermore,
foreground material would be polarized due to the magnetic fields 
at the Galactic center, and therefore polarization variations of
stars would  
provide a test of the hypothesis that the relative motion of streamers and stars
are responsible for stellar intensity variations.

\subsubsection{Variability of Stars Near Closest Approach}

\begin{figure*}
\epsscale{0.7}
\plotone{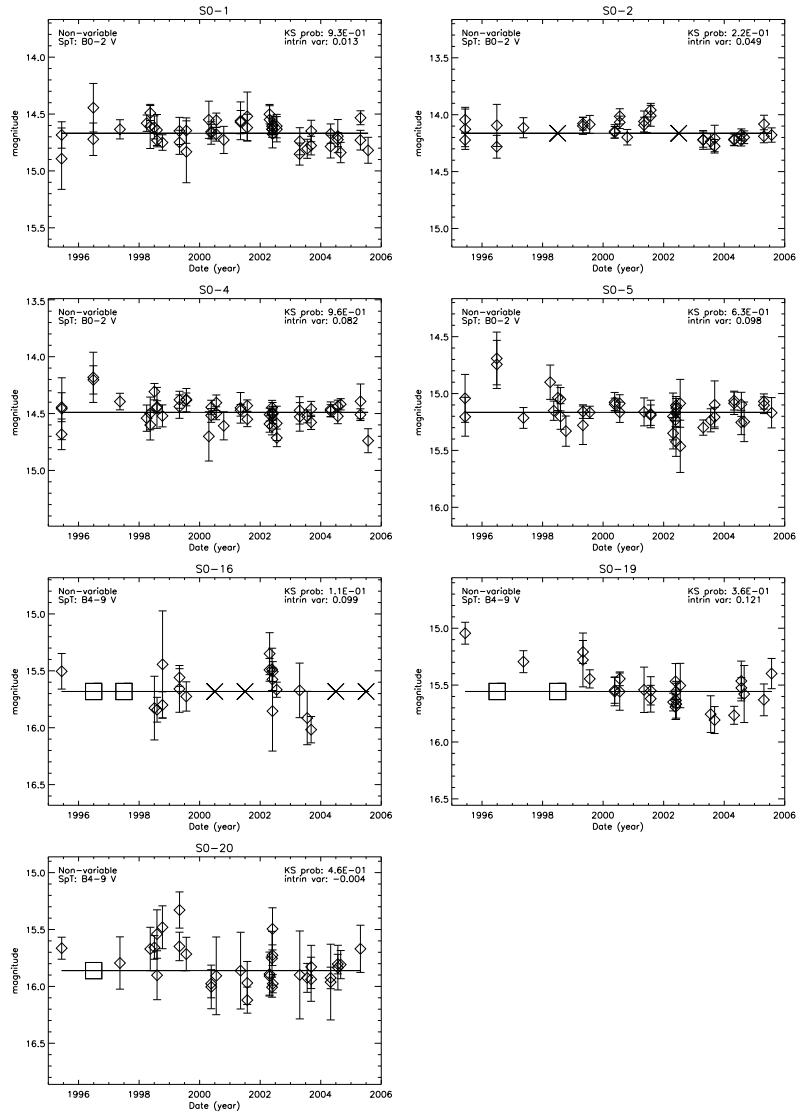}
\caption{Light curves of 7 central arcsecond sources that have known 3-dimensional orbits. 
The X's mark observations that are rejected due to confusion with other sources, and the boxes are areas where the 
nightly images have missing measurements due to insufficient image sensitivity 
and correspond to detections in monthly averaged images. None of these stars are variable,
which is evidence against the presence of a dusty disk around Sgr A$^{*}$.
The given intrinsic variance is expressed in magnitudes.
\label{lcorbit}}
\epsscale{1.0}
\end{figure*}

We detect no variability in the 7 central arcsecond sources that
have known 3-dimensional orbits 
(S0-1, S0-2, S0-4, S0-5, S0-16, S0-19, S0-20)
(see Figure \ref{lcorbit}). 
The three fainter stars (S0-16, S0-19, S0-20) have missing measurements that are 
due to insufficient image sensitivity, although they are detected
in higher signal-to-noise images made from  multiple nights of data \citep{ghez05a}.
The photometry of these stars constrains the properties of a cold, 
geometrically-thin inactive accretion disk around Sgr A$^{*}$, since in the presence 
of such a disk we would expect to see nearby stars significantly
flaring in the NIR as they passed through and interacted with the disk,
and eclipsed at other times.
When a nearby star approaches such a disk we would 
see enhanced NIR flux from reprocessed UV and optical starlight
incident on the disk (which we call a flare). 
Also, we would expect the disk to eclipse the star, reducing the flux from the star
in varying amounts depending on the properties of the disk. The time-scales
vary based on the geometry of the disk but are on the order of a year and months 
for the eclipses and flares, respectively \citep{nay03, cua03}. 
An optically thin disk may not fully eclipse 
stars, and gaps in our observations would allow different geometries of the disk 
to account for any one star not showing eclipses or flares as was calculated
for S0-2 \citep{cua03}. 
However, with the ensemble of stars that we have monitored, 
the effects of a disk with any orientation should be evident.  
We constrain the NIR optical depth to be $\lesssim0.1$. If we assume
the NIR standard interstellar dust opacity at 2.2$\micron$ where the
dust extinction is approximately $\sim5\times10^{-22} cm^{2}$ per hydrogen atom
(see Fig. 2 in \citealt{vosh03}), we find the column density of the disc to be
$\sim2\times10^{20} cm^{-2}$. However,
 dust grains may be larger in size or non-existant in the disc \citep{cua03, nay04}, and 
such constraints should therefore be approached cautiously.
Regardless, the lack of observed flares or eclipses in the 7 central arcsecond sources that
have known 3-dimensional orbits puts such severe constrains on the density and 
size of any possible disc disk around Sgr A$^{*}$ that such a disk is unlikely to exist.

\vspace{0.1in}

\section{Summary}

We use ten years of diffraction-limited K-band 
speckle data to determine the photometric
stellar variability in the central 5$\tt''$ $\times$ 5$\tt''$ of our Galaxy. 
Within this study's limiting magnitude of $m_{K} < 16$ mag,
we find 15 K-band variable stars out of 131 well-sampled stars.
Among 46 stars brighter than $m_{K} < 14$ mag with 
uniform photometric uncertainties, there are 10 variable stars,
suggesting a minimum variable star frequency of 23\%.
We find one periodic star, IRS 16SW, with a 
period of P=$19.448 \pm 0.002$ days, 
in agreement with \citet{ott99}, \citet{depoy04}, and \citet{mar06}. 
Our data are consistent with an eclipsing binary and
show a rise-time that is $\sim1.6$ times longer than the fall time
and we suggest that the asymmetric light curve
results from tidal deformations of the two stars in the presence of asynchronous rotation.
We expect to see on the order of one eclipsing binary in our sample
for conditions similar to the rest of the Galaxy,
suggesting that the frequency of eclipsing binaries is not
significantly increased at the 
Galactic center over the local neighborhood. 
We identify IRS 29N as a wind colliding binary based on its light curve
and spectral classification. This rare object warrants further investigation
to confirm its binary nature.
None of the IRS 16 stars shows the classic eruptive events of LBVs, although
our time baseline is too short to rule them out as LBVs. 
Among the remaining 
variable early-type stars in our sample, 3 exhibit
large variations on time-scales of a year, which are either due to 
obscuration from mass loss events or from line of sight extinction variations.
Three more stars in our sample exhibit long term variations of $\sim$5-10 yrs
probably due to line of sight extinction variations due to high density streamers.
Seven stars in the central arcsecond do not show photometric variations 
indicative of a cold geometrically thin inactive accretion disk which puts such
severe constraints on the density and size of any possible disk around Sgr A$^{*}$
that such a disk is unlikely to exist.


\acknowledgements

The authors thank Eric Becklin, Jill Naiman and Andrea Stolte for helpful 
conversations and useful comments on the manuscript
and an anonymous referee for a helpful review.
Support for this work was provided by NSF grant AST-0406816.
The W. M. Keck Observatory
is operated as a scientific partnership among the California Institute 
of Technology, the University of California and the National Aeronautics and 
Space Administration.  The Observatory was made possible by the generous 
financial support of the W. M. Keck Foundation.  The authors wish to recognize and 
acknowledge the very significant cultural role and reverence that the summit 
of Mauna Kea has always had within the indigenous Hawaiian community.  We are 
most fortunate to have the opportunity to conduct observations from this 
mountain.  

\pagebreak

\end{document}